\documentclass[12pt,dvipsnames]{iopart}
 \expandafter\let\csname equation*\endcsname\relax 
 \expandafter\let\csname endequation*\endcsname\relax 
\usepackage{amsmath,amsfonts,amssymb}

\pdfoutput=1
\usepackage{graphicx}
\usepackage{color}
\usepackage{appendix}

\begin{document}

\newcommand{\R}{\mathbb{R}}
\newcommand{\C}{\mathbb{C}}
\newcommand{\Z}{\mathbb{Z}}
\newcommand{\noi}{\noindent}
\newcommand{\goto}{\rightarrow}
\newcommand{\la}{\langle}
\newcommand{\ra}{\rangle}
\newcommand{\skp}{\vspace{\baselineskip}}

\newcommand{\rhotilde}{\tilde{\rho}}
\newcommand{\rhobar}{\bar{\rho}}
\newcommand{\rhoeq}{\rho_{eq}}
\renewcommand{\hbar}{\bar{h}}
\newcommand{\htilde}{\tilde{h}}
\newcommand{\sbar}{\bar{s}}
\newcommand{\stilde}{\tilde{s}}
\newcommand{\lambdadot}{\dot{\lambda}}
\newcommand{\lambdahat}{\hat{\lambda}}
\newcommand{\xidot}{\dot{\xi}}
\newcommand{\xihat}{\hat{\xi}}
\newcommand{\ahat}{\hat{a}}
\newcommand{\adot}{\dot{a}}
\newcommand{\bdot}{\dot{b}}
\newcommand{\xdot}{\dot{x}}
\newcommand{\ydot}{\dot{y}}
\newcommand{\zdot}{\dot{z}}
\newcommand{\Mhat}{\hat{M}}
\newcommand{\Ghat}{\hat{G}}
\newcommand{\Khat}{\hat{K}}
\newcommand{\Psihat}{\hat{\Psi}}
\newcommand{\wtilde}{\tilde{w}}
\newcommand{\half}{\frac{1}{2}}

\renewcommand{\d}{\partial}
\renewcommand{\L}{{\cal L}}
\renewcommand{\H}{{\cal H}}

\title{{\bf Coarse-graining two-dimensional turbulence via dynamical optimization} }
\author{ Bruce Turkington, Qian-Yong Chen  and  Simon Thalabard \\
 \small Department of Mathematics and Statistics  \\
  \small University of Massachusetts Amherst    }

\date{\small September 2015}

\normalsize
\begin{abstract}  \noindent
A model reduction technique based on an optimization principle is employed to coarse-grain  
inviscid, incompressible fluid dynamics in two dimensions.  In this reduction the spectrally-truncated
vorticity equation defines the microdynamics, while the macroscopic state space consists of 
quasi-equilibrium trial probability densities on the microscopic phase space, which are parameterized 
by the means and variances of the low modes of the vorticity.   A macroscopic path therefore represents 
a coarse-grained approximation to the evolution of a nonequilibrium ensemble of microscopic solutions.   
Closure in terms of the vector of resolved variables, namely, the means and variances of the low modes, 
is achieved by minimizing over all feasible paths the time integral of their mean-squared residual with 
respect to the Liouville equation.  The equations governing the optimal path are deduced from Hamilton-Jacobi 
theory.   The coarse-grained dynamics derived by this optimization technique contains a scale-dependent 
eddy viscosity, modified nonlinear interactions between the low mode means, and a nonlinear coupling 
between the mean and variance of each low mode.  The predictive skill of this optimal closure is validated 
quantitatively by comparing it against direct numerical simulations.   These tests show that good agreement 
is achieved without adjusting any closure parameters.    
  \end{abstract}

\noi {\it Key Words and Phrases: 
model reduction,   turbulence closure, nonequilibrium statistical mechanics, Hamilton-Jacobi theory.}
 \maketitle
\section{Introduction}
The phenomenology of two-dimensional turbulence has been extensively studied in many
numerical simulations and physical experiments \cite{KM,Lesieur,Tabeling,BE}.      
At high Reynolds number, free-decaying
turbulence in two dimensions exhibits both the emergence of coherent structures on large
scales and the development of vorticity fluctuations on small scales.   This generic behavior is
consistent with the conceptual picture of the dual cascade for forced, two-dimensional turbulence,
in which energy fluxes toward low wavenumbers while enstrophy  fluxes toward
high wavenumbers.    In either the free-decaying or forced scenarios,
there is significant dynamical interaction between the large and small scales ---  
coherent vortices emerge out of a background of vorticity fluctuations, while vortex coalescence 
expels filamentary vorticity into the vicinity.   As is 
well known, this typical behavior of two-dimensional flow is shared by the important asymptotic
models of geophysical fluid dynamics, notably, quasi-geostrophic flow \cite{Leith,Salmon,MW,Turk-lesHouches,BV}.    
Indeed, the geophysical equations are perhaps even more interesting from this point of view, 
since  their Reynolds numbers are huge, 
and  the effects of a nonzero Coriolis gradient or a nontrivial bottom topography 
introduces spatial inhomogeneities into the dynamics and consequent organization of the mean flow.

 From the perspective of both theoretical understanding and practical applications, therefore, it is highly desirable to develop a methodology for modeling the evolution of the coherent part of a two-dimensional, or quasi-geostrophic, flow without resorting to full simulation of all its active scales of motion.     Moreover, greater insight into behavior of such a complex system is gained from ensemble-averaged, coarse-grained descriptions than from the full integration of its chaotic dynamics \cite{GKS}.   Briefly put, one seeks a robust closure model for the low wavenumber behavior.   Of course, turbulence closures have a long history, beginning with Reynolds' identification of the effective stress due to fluctuations, followed by Boussinesq's concept of an eddy viscosity and Prandtl's mixing length \cite{MY1}. 
These semi-empirical models of the effect of the small-scale eddies on the large-scale flow have 
stimulated numerous efforts to
derive them coarse-grained models from the underlying fluid dynamics \cite{MY2,Orzsag,Lesieur,McComb}.   This historical development has been dominated by a single conceptual viewpoint, in which closure is achieved by terminating  the hierarchy of equations for the successively higher moments generated by the quadratic nonlinearity of hydrodynamics.    
Even though many useful closures have been developed and deployed,  a  systematic and justified model having a wide range of validity has not emerged from this line of reasoning.    Generally speaking, 
closure hypotheses of this kind have relied on formal
perturbation arguments or adjustments to the derived equations for moments, such as eddy damping, with the result that they
 have recourse to empirically fit parameters or heuristic arguments.     Moreover, most such investigations have focused on 
 homogeneous turbulence, even though inhomogeneity is a key feature of two-dimensional flows that may be included in
 statistical closures \cite{FO}.   Most of the work in this direction nonetheless reinforces a generalized notion of
eddy viscosity, which may vary with scale and may depend nonlinearly on various 
resolved quantities, such as the mean flow or turbulent kinetic energy \cite{Kraichnan}.           

In this paper we develop a new approach to turbulence closure in the 
context of  two-dimensional, incompressible,
inviscid fluid flow.  Our goal is to derive closed reduced equations for 
an appropriately coarse-grained vorticity field from the underlying fluid dynamics.  
To do so we combine some basic principles in nonequilibrium statistical mechanics with a new model reduction technique
 developed by one of the authors  in a general setting  of finite-dimensional Hamiltonian systems  \cite{Turk}.  
 For the underlying microscopic dynamics we adopt a spectral truncation of ideal vorticity transport equation; specifically,
 the Fourier-Galerkin truncation of ideal flow 
on a two-dimensional torus,  $x= (x_1,x_2) \in [0, 2\pi)^2$, using modal wavevectors 
$k= (k_1,k_2)$  with $ k_1, k_2 = 0 , \pm 1, \ldots , \pm n$, for $n \gg 1$.     
 In this sense our work is conceptually aligned with recent 
 work on the inviscid Burgers equation \cite{MT1,MT2,AKM} and three-dimensional Euler dynamics \cite{CBDB,KMBP},
 which investigate the turbulent behavior and thermalization of truncated conservative systems.      
 The coarse-grained  description retains only the low wavenumber modes, having
 $-m \le k_1, k_2 \le m$, for a fixed $m \ll n$,  and reduction is achieved by imposing a quasi-equilibrium statistical model,
in which the means and variances of the  low modes of are the resolved variables, and  
the unresolved modes  have equilibrium statistics determined by energy and enstrophy conservation.     
  
Rather than proceeding along the traditional lines of closing a hierarchy of moments,
our reduction method characterizes that coarse-grained macrodynamics which is optimally compatible with the 
fully-resolved  microdynamics.   Optimality is quantified 
by minimizing a cost functional over paths of macrostates, the cost being   
a net information loss rate incurred by reduction from the full deterministic dynamics
to the reduced statistical model.      The closed reduced equations derived from this optimization principle have a
generic thermodynamical structure, in which the reversible and irreversible parts of the reduced 
equations are separately identified, and the optimal cost of reduction is related to  entropy
production \cite{Ottinger}.   These properties give the optimal closure  
a clear physical interpretation and a natural mathematical structure.

The choice of quasi-equilibrium probability densities in the optimization principle makes the systematic derivation of an optimal  closure feasible.  
With respect to these statistical states  all modes are uncorrelated and Gaussian,
properties shared by the well-studied canonical statistical equilibria for the truncated two-dimensional Euler equations 
\cite{KM}.    
Statistical models based on these distributions and their refinements have been shown to describe 
the long-time, large-scale organization of two-dimensional, and especially quasi-geostrophic, turbulence
\cite{Salmon,MW,Turk-lesHouches,BV}.                
Besides tractability, the justification for using quasi-equilibrium densities
relies on the rapid relaxation of the unresolved modes to equilibrium, and the near-normality of the resolved modes.  
Since simulations and observations of two-dimensional turbulence show some 
non-normality in the higher modes of vorticity as well as no sharp separation in time scales between a
set of low modes and the remaining higher modes, we necessarily introduce some
error in our closure by choosing these convenient trial probability densities.     
Our optimization principle produces  the best closure that can be obtained under these simplifications.       

The present paper is closely related to recent work by Kleeman and one of the authors on the 
truncated inviscid Burgers equation \cite{KT1},  a prototype
having a hydrodynamical nonlinearity and exhibiting chaotic and mixing dynamics.   
The general model reduction technique developed in \cite{Turk} is applied to the low modes of this system,
and a closure in terms of their means is obtained.   The reduced model predicts that the
 decay rate of the resolved mode with wavenumber $k$ scales  as  $|k|$, a prediction that  
matches the scaling observed in direct numerical simulations.   Moreover, the reduced equations include
modifications to the quadratic nonlinearity in the Burgers equation involving a 
nonlocal bilinear operator acting on the mean resolved modes.   

In the context of two-dimensional turbulence, our optimal closure predicts that the rates of the decay to equilibrium 
of the means and variance perturbations scale with $ \,  |k| \, \sqrt{ \log  |k|} \, $, where   
 $k = (k_1,k_2)$ is the wavevector and $ |k| = \sqrt{k_1^2+k_2^2} $.   
Thus, within the quasi-equilibrium model in which the 
the unresolved modes have equilibrium statistics, the dissipative mechanism acting on the coarse-grained flow is found to be
a particular eddy viscosity.       
In addition, the optimal coarse-grained dynamics has two novel nonlinearities: 
(1)  modifications to the interactions between resolved modes resulting from 
the collective effect of  interactions via unresolved modes; 
and (2) a dynamical coupling between the mean and variance of each mode.     
Most significantly, this optimal closure is found to be universal and intrinsic in the sense that it
contains no adjustable parameters.

Our dynamically optimal Gaussian closure is developed in Sections 2 through 5.   The quasi-equilibrium
statistical model is defined in Section 2, and then in Section 3 the optimization principle over paths is 
articulated.    Section 4 explains how Hamilton-Jacobi theory produces the equations governing optimal paths.
Section 5 completes the derivation of the closure by solving the Hamilton-Jacobi
equation approximately using a near-equilibrium perturbation analysis.       
This development pertains to arbitrary truncation wavenumbers, $m$ and $n$,
for the macroscopic and microscopic descriptions, respectively, provided that $m \ll n$.   In Section 6 the model
equations are validated numerically against  direct simulations of ensembles propagated by the 
fully-resolved dynamics.   To ensure statistical convergence, this comparison is conducted on very large ensembles of a 
modestly truncated system, for which $m=4$ and $n = 32$;  this coarse-graining projects $2112$ degrees
of freedom onto $40$ resolved modes.   The magnitude of the relative error incurred by the optimal closure, and its dependence
on the amplitude of the nonequilibrium initial state, are revealed at this resolution.    
Benchmarking the model against direct statistical solutions on resolutions typical in fluid dynamics, 
where $m$ and $n$ are much larger,  is not attempted here due to the computational cost.   Nonetheless, our test results
suggest that the predictive skill of the optimal closure would be comparable at high resolution.

\section{Dynamics and statistics}

We study the dynamics of ideal flow in $\R^2$, for which the continuum governing
equations are 
\begin{equation}  \label{2deuler}
\frac{\d u}{\d t}  \, + \, u \cdot \nabla u \, = \, - \nabla p \, , \hspace{1cm}
\nabla \cdot u \, = \, 0 \, .     
\end{equation}
This dynamics is most conveniently represented in terms of the vorticity field, $\zeta=\zeta(x,t)$, 
and the streamfunction $\psi = \psi(x,t)$, for which
\[
\zeta =  \frac{\d u_2}{\d x_1} \, - \,  \frac{\d u_1}{\d x_2}  \, ,  \hspace{1cm}  
u = \left( \, \frac{\d \psi}{\d x_1} \, , \, - \frac{\d \psi}{\d x_2}  \, \right) \, .      
\]
Then (\ref{2deuler}) is equivalent to the pair of scalar equations
\begin{equation}  \label{vort-stream}
\frac{\d \zeta}{\d t}  \, + \, [ \, \zeta , \psi \, ]  \, = \, 0 \, , \hspace{1cm}
  -  \triangle \psi \, = \, \zeta \, ,   
\end{equation}  
where $[\phi,\psi] = \d \phi / \d x_1 \d \psi / \d x_2 \, - \,  \d \phi / \d x_2 \d \psi / \d x_1 $ denotes
the Poisson bracket of scalar fields on $\R^2$.     

In our investigations we consider doubly-periodic boundary conditions in the
space variable $x=(x_1,x_2)$.    Normalizing the periods we assume that $\zeta$ and $\psi$
are $2 \pi$-periodic in $x_1$ and $x_2$, and  
we write $x \in T^2 = [0, 2 \pi)^2$, the two-dimensional torus.      
Under these conditions it is necessary that 
\[
\int_{T^2} \zeta \, dx \, = \, \int_{\d T^2} u \cdot  dx \, = \,  0 \, .    
\]

We seek a coarse-graining of the dynamics (\ref{2deuler}), in which 
the small scales of motion, or the fine-grained fluctuations in vorticity, are represented 
by a dynamically consistent, statistical model.    The methods of statistical mechanics that we shall use are
applicable to dynamical systems with a large, but finite, number of degrees of freedom. 
We therefore replace the continuum dynamics (\ref{2deuler}) by its Fourier-Galerkin truncation.
Any doubly-periodic solution  $\zeta$ of (\ref{vort-stream}) 
is represented in terms of its complex Fourier coefficients 
\begin{equation}    \label{fourier}
z_k \, = \,z_k(t)  \, = \,  \frac{1}{(2 \pi)^2} \int_{T^2} \zeta(x,t) \, e^{- i k \cdot x} \,  dx   \,  \in \C \, , 
\end{equation}
for wavevectors $k = (k_1,k_2) \in \Z^2$, with $z_{(0,0)} =0$.  
 Truncation is imposed for $ |k_1|,| k_2| \le n$, for a fixed
cut-off wavenumber $n \gg 1$.    The vorticity field is then 
\[
\zeta(x,t) \, = \, \sum_{k \in K_n} z_k(t) \, e^{ i k \cdot x} \, , 
\]     
where this sum (and others to follow) extends over the lattice of nonzero wavevectors 
\[
K_n \doteq \{ \,   k = (k_1,k_2)    \in \Z^2 \, : \,    k_1, k_2 = 0, \pm 1 , \ldots , \pm n \, ,  \;\;
          k  \neq (0,0) \, \}   \, .   
\]
Since $\zeta$ is real-valued, $z_{-k} = z_k^*$, [star denotes  complex conjugate].  Hence
we may consider the microstate $z$ to be a point in the phase space  
$ \Gamma_n = \C^{d_n}$, having $d_n= [(2n+1)^2-1]/2= 2n(n+1)$ degrees of freedom.
The independent components of $z \in \Gamma_n$ correspond to wavevectors belonging
to half of the lattice, $K_n$, which may be partitioned into equal halves, 
$K_n = K_n^{+} \cup K_n^{-}$, with $K_n^{-} = - K_n^{+}$ in many ways.   
For convenience  we shall continue to write
 $z = ( \, z_k \, )$ for $ k \in K_n$, for a microstate, understanding that only half of
these variables are independent.    
 
 By applying the identity
 \(   
 [ \,  e^{ i p \cdot x}  \, , \,  e^{ i q  \cdot x}   \, ]  \, = \, - ( p \times q ) \,  e^{ i (p+q) \cdot x} \, , 
 \)
where  $\, p \times q \doteq p_1 q_2 - p_2 q_1 \, , $
for any wavevectors $p , q \in K_n$,  and using that the 
$k$-th Fourier coefficient of $\psi$ is $|k|^{-2} z_k$,  the  continuum equations
(\ref{vort-stream})  project onto the identities 
\[  
\sum_{ k \in K_n}  \left[ \frac{ d z_k}{dt} \, - \, 
    \sum_{p+q=k} \frac{p \times q}{|q|^2 } \, z_p z_q \,   \right]  e^{ i k \cdot x}  \; = \, 0 \, ,
    \hspace{1cm} \mbox{ for all } \; k \in K_n \, . 
\]
The spectrally truncated microdynamics on the phase space $\Gamma_n$  is therefore
governed by the system of ordinary differential equations 
\begin{equation}  \label{microdynamics}
\frac{ d z_k}{dt} \, + \,    \sum_{p+q=k }  c (p,q) \, z_p z_q \, = \, 0 
      \hspace{1cm}    ( \,   k , p , q \in K_n \, ) \, , 
\end{equation}
where we introduce the symmetrized interaction coefficients
\begin{equation}   \label{interaction-coeff}
c(p,q) \, \doteq \,  \frac{1}{2} (p \times q) \left(  \frac{1}{|p|^2 } -  \frac{1}{|q|^2 }\right) 
 \, = \, c(q,p)  \hspace{2cm}
    ( \, p , q \in K_n \, ) \, .  
\end{equation}  
These  coefficients are real,  and $c(-p,q) = c(p,-q) = - c(p,q)$.    
The symmetrized form of the microdynamics is convenient
in many of the subsequent calculations and hence will be employed throughout the sequel.

The dynamics (\ref{microdynamics}) conserves both energy and enstrophy, respectively,
\begin{eqnarray}  \label{energy}
H(z)  & = & \frac{1}{2} \sum_{k \in K_n} |k|^{-2} \, |z_k|^2 \, ,   \\  \label{enstrophy}
Z(z)  & = &   \frac{1}{2} \sum_{k \in K_n}  \, |z_k|^2 \, .
\end{eqnarray}   
The invariance of $H$ and $Z$ are consequences, respectively, of the two identities 
\begin{equation}  \label{interaction-identities}
c(p,q)|k|^{-2} = c(q,k)|p|^{-2} + c(k,p)|q|^{-2}  \, ,  \;\;\;\;\;\;   c(p,q) = c(q,k) + c(k,p) \, .
\end{equation}
Though we will refer to  $Z$ as the enstrophy, it is actually the {\em quadratic} enstrophy.   
For the continuum dynamics there are infinitely many independent enstrophy integrals,
$Z_f = \int f(\zeta) \, dx $, for any (regular) real function $f$ on the (invariant)
range of the vorticity field.    But only the quadratic entrophy (\ref{enstrophy}) survives 
Fourier modal truncation as an exact invariant \cite{KM,Lesieur,MW}.     

The Liouville theorem holds for the (noncanonical Hamiltonian) dynamics (\ref{microdynamics}), 
so that  phase volume $dz = \prod_k dx_k dy_k$  
on $\Gamma_n$, writing $z_k=x_k + i y_k$,  is dynamically invariant under the phase flow \cite{KM,Lesieur}.    
Among the equilibrium statistical distributions on $\Gamma_n$  determined by the 
energy and enstrophy invariants is the canonical probability density 
\begin{equation}  \label{equil-distr}
\rho_{eq}(z)   \, = \, \exp ( - \beta Z - \theta H - \Phi(\beta,\theta) \, ) \, , 
\end{equation}
where 
\[
\Phi(\beta,\theta) \, = \, \log \int_{\Gamma_n}  \exp ( - \beta Z - \theta H ) \, dz \, .   
\]
The  parameters $\beta$ and $\theta$ conjugate to mean enstrophy and energy, respectively, 
are required to satisfy $\beta >0$ and 
$\theta > - \beta$, so that $\beta + \theta |k|^{-2}$ is positive for all $k \in K_n$.     For these       
admissible values, the Fourier coefficients $z_k$ are  complex Gaussian
random variables with first and (isotropic) second moments 
\begin{equation}  \label{equi-gaussian}
\la z_k \ra_{eq} = 0 \, , \;\;\;\;\;\;   
    \la |z_k|^2 \ra_{eq} = \frac{1}{ \beta_k}  \;\;\;\;\;\;   \mbox{ where} \;\;\;   \beta_k \doteq \beta + \theta |k|^{-2} .
\end{equation} 
The invariance of $\rho_{eq}$ under the microdynamics (\ref{microdynamics}) is a consequence of
the identity
\begin{equation}   \label{c-beta-identity}
c(p,q)\beta_k = c(q,k)\beta_p +  c(k,p)\beta_q 
\end{equation}
which is immediately implied by (\ref{interaction-identities}).  

In our nonequilibrium statistical closure the equilibrium density (\ref{equil-distr}) furnishes a
convenient  and natural reference distribution for a nonequilibrium theory.   
This distribution is not intended to be a realistic representation of developed two-dimensional
Navier-Stokes turbulence in either a freely-decaying or forced scenario.   Far-from-equilibrium 
phenomena arising from viscous dissipation, such as cascades, require much more
complicated statistical descriptions.      In the present paper, we endeavor to coarse-grain
the inviscid Euler dynamics itself, and hence we construct our nonequilibrium distributions 
from $\rho_{eq}$, rather than introducing nonequilibrium behavior
into the unresolved fluctuations.    A more elaborate theory might derive a coarse-grained effective
equations for the low modes with respect to a background of vorticity fluctuations that have a
nonequilibrium spectrum.  Such a theory is a worthwhile goal for future research.     

The nonequilibrium statistics of the full dynamics (\ref{microdynamics}) 
is exactly represented by the propagation of probability density $\rho(z,t)$ with respect to $dz$ 
under the Liouville equation, 
\begin{equation}  \label{liouville}
0 \, = \, \frac{\d \rho}{\d t} + \sum_{k \in K_n^+}  
    \frac{dx_k}{dt}  \frac{\d \rho}{\d x_k} + \frac{dy_k}{dt} \frac{\d \rho}{\d y_k}  \, = \, 
   \frac{\d \rho}{\d t} + \sum_{k \in K_n}
   \frac{dz_k}{dt}   \frac{\d \rho}{\d z_k}  \, .     
\end{equation}
As is universally recognized throughout statistical mechanics,  the complete determination of 
the evolving density $\rho(z,t)$ is not feasible or desirable for high-dimensional systems  
\cite{Balescu,Balian,Tuckerman,Zwanzig}.   Indeed, the full complexity of $\rho(z,t)$ is not needed 
 to approximate the statistics of coarse-grained observables.   
In this light we contract the solution space of (\ref{liouville})
to a space of approximations, which we call trial probability densities and denote by $\rhotilde(z,t)$.  
Our reduction employs the Gaussian trial densities
\begin{equation}  \label{trial-densities}
\rhotilde(z;a,b) \, = \, \prod_{k \in K_n^+}  \frac{b_k}{\pi} 
e^{ - b_k \, | \, z_k  - a_k \, |^2 }  \; ,  
\end{equation}  
which are parameterized by $a_k \in \C$, $b_k \in (0,\infty)$.   
That is, we impose parametric statistical model 
with independent Gaussian modes, having means and variances 
\begin{equation}    \label{means-variances}
\la \, z_k \, | \, \rhotilde(a,b) \, \ra \, = \,a_k \, , \;\;\;\;\;\;\;\;
\la \, | \, z_k - a_k \, |^2  \, | \, \rhotilde(a,b) \, \ra \, = \, \frac{1}{b_k} \, .   
\end{equation}     
Throughout the paper the expectation of any observable $F$ on the phase space $\Gamma_n$ 
with respect to a probability density $\rho$ is denoted by 
\(
\la F \, | \, \rho \, \ra  = \int_{\Gamma_n} F(z) \, \rho(z) \, dz \, .  
\)     
 For convenience in all subsequent expansions over modes, we extend 
the statistical parameters $a_k, b_k$ to the entire lattice of wavevectors $K_n$, by setting
$a_{-k} = a_k^*$, $b_{-k} = b_k$, for  $k \in K_n^+$.   

We coarse-grain the statistical dynamics by fixing a low wavenumber cutoff, $m \ll n$,
and setting the parameters $a_k$ and $b_k$ to equilibrium values for $\max \{ |k_1|, |k_2| \}  > m$.   
That is, in the trial densities (\ref{trial-densities}) we put  
$a_k=0$ and $b_k = \beta_k = \beta + \theta |k|^{-2} $ for wavevectors $k \in K_n \backslash K_m$
of the unresolved modes, and we adopt the parameters $a_k$ and $b_k$ 
for wavevectors $k \in K_m^+$ of the low modes as the resolved variables in our reduced model.     
The equilibrium density, $\rho_{eq}$,  itself belongs to the statistical model and corresponds to 
$a_k=0, \, b_k = \beta_k$ for all $k \in K_n$.    The trial densities (\ref{trial-densities}) 
are quasi-equilibrium statistical states in which the low modes are disequilibriated,
while the remaining higher modes are equilibriated.      
These densities may be realized as
maximum entropy states,  in the sense that  $\rhotilde(z;a,b)$ is the unique solution of 
the constrained maximization problem: 
\[
\max_{\rho} \,  -  \la \, \log \rho   \, | \, \rho \, \ra \;\;\;\;\; \mbox{ over } \;\;
\la \, z_k \, | \, \rho \, \ra \, = \,a_k \, , \;\;
\la \, | \, z_k - a_k \, |^2  \, | \, \rho \, \ra \, = \, \frac{1}{b_k} \,  .
\]
Conceptually, the use of quasi-equilibrium states as trial densities rests on the notion 
that the unresolved modes decorrelate and relax toward equilibrium more rapidly than the resolved modes \cite{Zubarev}.  
Nonetheless, the division between 
the resolved  and unresolved modes is arbitrary in the context of turbulence closure,  as it 
does not rest on some intrinsic separation of time scales.       
Consequently, we expect that our reduced model
will entail some irreducible error arising from memory effects as well as the errors associated with
the Gaussian approximation \cite{Zwanzig,GKS}.

\section{Optimal paths}   

Our interest centers on predicting the evolution of the means and
variances of the lowest modes, that is, the statistical parameter vector,
$(a_k(t),b_k(t)) $, for $k \in K_m^+$,  via a closed dynamics in these resolved variables alone.   
We do so by considering all admissible paths $(a(t),b(t))$ in the statistical parameter 
space, $\C^{d_m} \times (0,\infty)^{d_m}$, where $d_m = 2m(m+1)$ is the cardinality
of the reduced half-lattice $K_m^+$ of resolved modes.   Since our attention is focused
on relaxation of a  nonequilibrium initial state toward equilibrium, we introduce a
cost functional designed to quantify the compatibility of a path of trial densities
$\rhotilde(z; a(t),b(t))$ to the underlying microdynamics, and 
we minimize the time integral of the cost function over $t \in [0,\infty)$.  
The construction of the appropriate cost functional is as follows.   

The fundamental statistic associated with an admissible path $(a(t),b(t))$ is
its residual with respect to the Liouville equation (\ref{liouville}), namely,  
\begin{equation}   \label{residual}
R \, = \,  \left( \,  \frac{\d }{\d t} + \sum_{k \in K_n}  \frac{d z_k}{dt} \frac{\d }{\d z_k} \, \right) 
                    \log \rhotilde(\cdot \, ; a(t),b(t))   
    \, = \, \sum_{k \in K_n}  \adot_k^* U_k + \frac{1}{2}  \bdot_k V_k
                    -     U_k^* \frac{d z_k}{dt} \, .
\end{equation}   
In this expression, $\adot_k = d a_k /dt$, $\bdot_k = d b_k /dt$, and 
\begin{eqnarray}   \label{score}
U_k &=& \frac{\d}{\d a_k^*} \log \rhotilde(\cdot; a,b) \,= \, b_k (z_k-a_k)   \\
V_k &=& \frac{\d}{\d b_k}  \log \rhotilde(\cdot; a,b) \,= \,  \frac{1}{b_k} - |z_k-a_k|^2   \, .                             
              \nonumber 
\end{eqnarray} 
$d z_k/dt$ is given by the microdynamics (\ref{microdynamics}).  
For $k \in K_m$, $U_k, V_k$ denote the score functions associated with the parametric statistical model \cite{CB}.  
For $ k \in K_n \backslash K_m$, we extend them to be $U_k=  \beta_k z_k$, 
$V_k=  1/\beta_k - |z_k|^2$, consistently with the extension $a_k=0$, $b_k=\beta_k$.   Also, $U_{-k} = U_k^*$ and
$V_{-k} = V_k$ for all $k \in K_n$.    The score functions satisfy  
$ \, \la U_k \, |  \, \rhotilde \ra  = \la V_k \, |  \, \rhotilde \ra = 0 \,$,
and 
\begin{equation}   \label{second-moments} 
\la U_{k} U_{k'}^* \, |  \, \rhotilde \ra = b_{k} \delta_{k, k'} \, , \;\;\;\;
\la V_{k} V_{k'} \, |  \, \rhotilde \ra = \frac{1}{b_{k}^2} [ \,  \delta_{k, k'} +  \delta_{k ,-k'}\, ] , \;\;\;\;
\la U_{k} V_{k'} \, |  \, \rhotilde \ra = \la U_{k}^* V_{k'} \, |  \, \rhotilde \ra =0 \, . 
\end{equation} 
We refer to the span of the score functions (\ref{score}) for $k \in K_m^+$ as the resolved subspace of $L^2(\Gamma_n, \rhotilde(a,b))$.  

Two interpretations of the  function $R=R(z;a,\adot,b,\bdot)$ are given in \cite{Turk}.
First, $R$ represents the rate of information loss due to reduction via the statistical model (\ref{trial-densities}), 
being the leading term in the local time expansion of the log-likelihood ratio between the evolved trial density 
and the exactly propagated density.    Second, $R$ satisfies the family of identities
\[
\frac{d}{dt}  \la \, F \, | \, \rhotilde(t) \, \ra \, - \,  \la \, \frac{dF}{dt} \, | \, \rhotilde(t) \, \ra \,
 = \,  \la \, FR \, | \, \rhotilde(t) \, \ra \, , 
\]  
for all observables $F$ with $\d F / \d t=0$; $dF/dt$ denotes the microscopic dynamics
applied to $F$.     
Since $\la R | \rhotilde(t) \ra = 0$, the covariance between the observable $F$ and the Liouville
residual $R$ quantifies the deficiency of the path of trial densities $\rhotilde(a(t),b(t))$
to propagate the expectation of $F$.   Together these properties motivates the our
choice of the  $L^2(\Gamma_n, \rhotilde(a,b))$ squared norm of $R$
as the cost function to measure the loss rate along paths in the reduced model.   
These concepts are more fully discussed in \cite{Turk,Kleeman}.

The projection of $R$ onto the resolved 
space and its complement, the unresolved subspace, defines the natural separation
of the reduced dynamics into its reversible and irreversible components \cite{Ottinger,Turk}.    
The orthogonal projection of any observable $F$ in $L^2(\Gamma_n, \rhotilde)$ with 
$\la F \, | \, \rhotilde \ra =0$ onto the resolved subspace spanned by $U,V $ is denoted by
$ P_{a,b} F $;
the complementary projection operator onto the unresolved subspace is $Q_{a,b} = I - P_{a,b}$.  
The separation of the Liouville residual, $R$, into its resolved and unresolved components
is facilitated by the fact that $R$ is a linear combination of the functions, $U_k, V_k$, and
the products,
$U_p U_q$, with $p+q \neq 0$,  and $U_p U_q U_k^*$, with $p+q=k $.   The
restrictions on the indexing wavevectors together with the independence of the score functions
implies that all these functions are mutually orthogonal (uncorrelated) in $L^2(\Gamma_n, \rhotilde)$.    
Thus, it suffices to exhibit $R$ as the following linear combination:  
\begin{eqnarray}    \label{expression-for-R}  
R&=& 
 \sum_{k \in K_n} \left[ \, \adot_k + \sum_{ p+q=k} c(p,q) \, a_p \, a_q \,  \right]^* U_k
                 \, + \, \frac{1}{2} \bdot_k V_k    \\ 
    & & - \,    \sum_{k \in K_n} a_k^* \sum_{ p+q=k} 
            \, \left[\,  \frac{c(q,k)}{b_q} +  \frac{c(k,p)}{b_p} \, \right] U_p U_q   \nonumber \\ & &
     +  \, \frac{1}{3}  \sum_{k \in K_n} \sum_{ p+q=k}   
       \left[ \, \frac{c(p,q)}{b_p b_q}  - \frac{c(q,k)}{ b_q b_k}  -  \frac{c(k,p)}{b_k b_p }  \right] U_pU_qU_k^*  \nonumber
\end{eqnarray}  
In these sums the wavevectors $p$ and $q$ run over the entire spectrum $ K_n$. 

The calculations leading from (\ref{residual}) to (\ref{expression-for-R})  are as follows.   
The two terms in (\ref{residual}) involving $\adot_k$ and $\bdot_k$ 
appear in (\ref{expression-for-R}) without manipulation.
The third term in (\ref{residual}) is calculated by substituting the governing equations
(\ref{microdynamics}) into it and then separating the terms which involve
single, double and triple products of the score functions;  namely,  
\begin{eqnarray}   \label{calc-R}    
 - \sum_{k \in K_n} U_k^* \frac{d z_k}{dt} & = & 
 \sum_{ k} \sum_{p+q=k}  c(p,q) \, 
      \left( \, a_p + \frac{1}{b_p} U_p \,   \right)   \left( \, a_q + \frac{1}{b_q} U_q \, \right)  U_k^*  \\                
           \nonumber   
       & = & S_{I} + S_{II} + S_{III} \, ,  
\end{eqnarray}   
where
 \begin{eqnarray*}
S_{I} =  \sum_{ k} \left[ \, \sum_{p+q=k}  c(p,q) \, a_p \,  a_q  \, \right] \, U_k^*   \,  = \, 
                  \sum_{ k} \left[ \, \sum_{p+q=k}  c(p,q) \, a_p \,  a_q  \, \right]^* \, U_k  \,  ,
 \end{eqnarray*}
 \begin{eqnarray*}
S_{II} =  \sum_{ k}  \, \sum_{p+q=k}  c(p,q) \, 
  \left[  \frac{a_p}{b_q} U_q \, + \,  \frac{a_q}{b_p} U_p \, \right] \, U_k^*   
   =  - \sum_{ k}  \, a_k^* \sum_{p+q=k}  \left[ \,  \frac{c(q,k)}{b_q}  + \frac{c(k,p)}{b_p}\, \right] U_p U_q    \, ,    
 \end{eqnarray*}   
 \begin{eqnarray*}         
S_{III}  =  \sum_{ k}  \, \sum_{p+q=k}  \, 
 \frac{c(p,q)}{ b_p b_q}  \, U_p U_q  U_k^*  
     =  \frac{1}{3} \sum_{k} \sum_{p+q=k} 
            \left[ \,  \frac{c(p,q)}{b_p b_q}  - \frac{c(q,k)}{b_q b_k }  -  \frac{c(k,p)}{ b_k b_p}  \, \right] U_p U_q U_k^*  \, .             
\end{eqnarray*}
$S_I$ is a linear combination of $U_k$ for $k \in K_m$ and $k \in K_n \backslash K_M$, 
and thus involves both resolved and unresolved components.    
The expression for $S_{II}$ follows from reindexing and then
symmetrization  over  wavevector indices.   
The expression for $S_{III}$ requires only symmetrization over the triply-indexed sum. 
  Both  $S_{II}$ and $S_{III}$ consist entirely
of unresolved components, since    $S_{II}$ is a linear combination of   products $U_p U_q$, and
 $S_{III}$ is a linear combination of the triple produces $U_p U_q U_k^*$, for which $p+q=k \neq 0$.

The cost function for our optimization principle is declared to be
\begin{equation}  \label{cost-fn-def}
\L(a, \adot,b, \bdot)  \, = \, \half  \la \, [ P_{a,b} R (a, \adot,b, \bdot)  ]^2  \, | \, \rhotilde (a,b) \, \ra
     +    \half \la \, [ W Q_{a,b} R (a, \adot,b, \bdot) ]^2  \, | \, \rhotilde(a,b) \, \ra  \, . 
\end{equation}
In the unresolved component of this expression we include a weight operator $W$, 
which is a self-adjoint, linear operator on $L^2(\C^n, \rhotilde)$ satisfying
$W Q = Q W$.    The unresolved subspace is
therefore invariant under $W$, and the effect of $W$ in (\ref{cost-fn-def}) is to assign weight to the
unresolved component of the residual $R$ relative to the resolved component.      
The weight operator contains all the primitive free parameters in our closure scheme.
These primitive weights may be viewed as encoding the influence of the unresolved 
modes on the resolved modes.  

The simplest choice is $W=1$ results in a cost function (\ref{cost-fn-def}) equal to
half the $L^2(\C^n, \rhotilde)$-norm squared of the residual $R$.   In  \cite{Kleeman} Kleeman advocates for
this universal choice  arguing that then the cost function  represents the
leading contribution to the relative entropy increment between an exactly propagated density and the
evolved trial density over a short time interval.  An attractive feature of this choice
is that the resulting optimal closure then involves no adjustable parameters.   

To test this information-theoretic argument, 
we consider a class of weight operators $W$,  and in Section 6, when quantifying the 
error between model predictions and direct numerical simulations of ensembles, we assess 
the choice $W=1$ against best-fit weights within the class.        
Since the unresolved subspace is spanned by products of the score variables $U_k$, 
it suffices to define $W$ on these products.   A natural choice of the weight operator $W$ 
is therefore
\begin{equation} \label{weight}
W (U_k) = \lambda_k U_k \, , \;\;\;\;
W ( U_p U_q ) = \mu_{p,q} U_p U_q \, , \;\;\;\; 
W ( U_p U_q U_k^* ) = \nu_{p,q} U_p U_q U_k^*   \, , 
\end{equation}
using weight factors $\lambda_k$,  $ \mu_{p,q}$ and  $\nu_{p,q}$ that are
arbitrary non-negative constants symmetric in their wavevector indices,
$p,q \in K_n$ with $p+q=k$.      For this class of weight operators the many
constants $\lambda_k$,  $ \mu_{p,q}$ and  $\nu_{p,q}$ are the primitive free parameters in the closure.    
We remark, however, that even when general weights are retained in the optimal closure theory, there are
far fewer adjustable parameters in the ensuing reduced equations, because only certain collective combinations
of the primitive weights enter into those equations.    
 
The evaluation of the cost function (\ref{cost-fn-def}) requires calculating moments up to
order 6.  The centered variables $U_k$ are Gaussian and uncorrelated,  
their second moments are given by (\ref{second-moments}), and 
all their odd moments vanish.    It suffices therefore to calculate 
\begin{eqnarray}   \label{higher-moments}  
\la \, U_{k_1} U_{k_2} U_{k'_1}^* U_{k'_2}^*\, | \, \rhotilde \, \ra & = &  b_{k_1}b_{k_2} 
        [ \,  \delta_{k_1 k'_1} \delta_{k_2 k'_2} + \delta_{k_1 k'_2}\delta_{k_2 k'_1} \, ]   \, , 
          \\ 
  \la \, U_{k_1} U_{k_2}U_{k_3} U_{k'_1}^* U_{k'_2}^*U_{k'_3}^*\, | \, \rhotilde \, \ra & = & 
      b_{k_1}b_{k_2}b_{k_3}
        [ \,  \delta_{k_1 k'_1} \delta_{k_2 k'_2} \delta_{k_3 k'_3} + \mbox{permutations} \, ]   \, .  \nonumber
\end{eqnarray}
Using the properties (\ref{second-moments},\ref{higher-moments}), 
we obtain the resolved and unresolved components of the cost function (\ref{cost-fn-def}), respectively, 
\begin{eqnarray}  
 \la \, [ P_{a,b} R  ]^2  \, | \, \rhotilde \, \ra &=&  
 \sum_{k \in K_m}   \left| \, \adot_k +  \sum_{ p+q =k} c(p,q) a_p a_q \right|^2 b_k
 \, + \, \frac{\bdot_k^2}{2 b_k^2}  \label{PRsq}  \\   \nonumber
     \la \, [ W Q_{a,b} R  ]^2  \, | \, \rhotilde \, \ra  &=&   
          \sum_{k \in K_n \backslash K_m}  \lambda_k^2
            \left| \, \sum_{p+q=k} c(p,q) a_p a_q \, \right|^2 \beta_k    \label{QRsq}     \\ 
  & &   + \,  2  \sum_{k \in K_m}  |a_k|^2 \sum_{ p+q=k} \mu_{p,q}^2 
  \frac{ \left[ \, c(q,k) b_p + c(k,p) b_q \, \right]^2 }{b_p b_q}    \\ 
  & &   + \, \frac{2}{3}  \sum_{k \in K_n} \sum_{p+q=k} \nu_{p,q}^2   
     \frac{ \left[   \, c(q,k) b_p + c(k,p) b_q - c(p,q) b_k  \right]^2}{b_p b_q b_k}
                         \,    . \nonumber 
\end{eqnarray}

The optimal closure is defined by minimizing  
the time integral of cost function over all macroscopic paths
connecting a given initial macrostate at time $t=0$ to equilibrium as $t \rightarrow \infty$;  namely,
\begin{equation}    \label{value-fn}
v(a^0,b^0) \, = \,   \min_{(a(0),b(0)) = (a^0,b^0)} \; \int_{0}^{+ \infty}  
        \L(a, \adot,b,\bdot)  \, dt \,  , 
 \end{equation}   
in which the admissible paths 
$(a(t),b(t))$, $0 \le t < + \infty \, $, in the configuration space of the
statistical model are constrained to start  at $(a^0,b^0)$ at time $t=0$. 
   
The value function $v(a^0,b^0)$ for the minimization problem (\ref{value-fn}) measures
the total lack-of-fit of the optimal path emanating from the initial conditions
$(a^0,b^0)$.  Finiteness of $v(a^0,b^0)$ ensures that the optimal path tends to equilibrium
as $t \rightarrow +\infty$;  at equilbrium $a_k=0$ and $b_k=\beta$ for all $k$ and hence the 
Liouville residual $R$ and its weighted mean square, $\L$, both vanish.    
As shown in the general theory paper \cite{Turk}, the value function  (\ref{value-fn}) is 
intimately related to the entropy production along the optimal path.

\section{Closure via Hamilton-Jacobi theory}

The closed reduced equations satisfied by the optimal paths for (\ref{value-fn}) 
are most effectively derived by invoking Hamilton-Jacobi theory of the calculus of variations \cite{Evans,GF,Lanczos,Sagan}.   
In a more compact form the cost-function (\ref{cost-fn-def}) is 
\begin{equation}   \label{cost-fn-compact}
\L(a, \adot,b,\bdot) \, = \, \sum_{k \in K_m^+}
        \left| \,   \adot_k  -  f_k(a) \, \right|^2 \, b_k \, + \,   \frac{\bdot_k^2}{ 2  b_k^2} 
            \; + \; w(a,b) \, ,
\end{equation}
introducing the shorthand notation
\begin{eqnarray}  
f_k(a) &  = &     \la \, \frac{d z_k}{dt}  \, | \, \rhotilde(a,b) \, \ra 
             \,   =  \,  -  \sum_{p+q = k} c(p,q) \, a_p a_q  \, ,  \label{f} \\                     
w(a,b) & = & \frac{1}{2}  \la \, [ W Q_{a,b} R  ]^2  \, | \, \rhotilde \, \ra \, ,  \label{w}
\end{eqnarray}  
where the last expression is given in (\ref{QRsq}).  
By analogy to classical mechanics, $\L$ in (\ref{cost-fn-compact}) may be considered a
Lagrangian for the reduced dynamics, in which  $- w(a,b)$ plays the role of the potential.    
But in making this analogy, it is essential to restrict the admissible trajectories to be only those which tend to 
equilibrium $(a,b) = (0,\beta)$ as $t \rightarrow +\infty$.      
 
 The  Legendre transform supplies the Hamiltonian $\H(a,\phi,b,\psi)$ conjugate to 
 $\L(a,\adot,b,\bdot)$.   The conjugate variables are 
 \begin{equation}  \label{conjugates}
\phi_k =  \frac{ \d \L }{\d \adot_k^*} \, = \, [ \, \adot_k - f_k(a) ] b_k \, , \;\;\;\;\;\;\;\;\;\;\;
\psi_k  = \frac{ \d \L }{\d \bdot_k} \, = \,  \frac{\bdot_k}{ b_k^2} \, ,  \\
\end{equation}
and the Hamiltonian function is 
\begin{eqnarray}  \label{hamiltonian}
\H(a,\phi,b,\psi) &=&  \sum_{k \in K_m^+} \phi_k ^* \adot_k  +   \adot_k^* \phi_k  \, + \, 
                                     \psi_k   \bdot_k  \,- \, \L  \\  
        &=&  \sum_{k \in K_m^+} \frac{ |\phi_k|^2 }{b_k}  + \phi_k^*f_k(a)    + f_k(a)^*\phi_k  
                 +  b_k^2 \frac{\psi_k^2}{2}   
                     \; - \; w(a,b) \, .  \nonumber 
\end{eqnarray}
In Appendix C these  expressions are shown to be
the appropriate complexifications of the real Legendre transform.    

The value function, $v(a,b)$, defined in  (\ref{value-fn})  
is analogous to an action integral, and hence it solves the
stationary Hamilton-Jacobi equation,
\begin{equation}  \label{HJ}
\H \left( a, \, - \frac{\d v}{\d a^*} ,\,  b, -  \frac{\d v}{\d b} \right) \, = \, 0 \, ;
\end{equation}
the minus signs in (\ref{HJ}) are due to imposing the 
constraints in (\ref{value-fn}) on the initial state $(a(0),b(0))$.  
The value function also satisfies the equilibrium conditions
\[
v(0,\beta) = 0 , \;\;\;\;\; \frac{ \d v}{\d a^*} (0,\beta) = 0 , \;\;\;\;\; \frac{ \d v}{\d b} (0,\beta) = 0 \, . 
\]
Hamilton-Jacobi theory dictates that along an optimal path $(a(t),b(t)) $
the conjugate variables are given by the canonical relations 
\begin{equation}   \label{conjugate-relations}
\phi_k(t) \, = \,   - \frac{\d v}{\d a_k^*}(a(t),b(t))     \, , \;\;\;\;\;\;\;\;   
\psi_k(t)  \, = \,   - \frac{\d v}{\d b_k}(a(t),b(t)) 
\end{equation}  
These relations then close the reduced dynamics along the optimal path
in terms of the value function; that is, (\ref{conjugates}) and
(\ref{conjugate-relations}) combine to form the closed system 
\begin{eqnarray}  \label{closed-reduced}  
 \frac{d a_k}{dt}  =  f_k(a)  \, - \, \frac{1}{b_k} \frac{\d v}{\d a_k^*}(a,b)  \, ,  \hspace{1cm} 
  \frac{d b_k}{dt}  =  - \,  b_k^2 \, \frac{\d v}{\d b_k}(a,b)       \, .  
\end{eqnarray}    
The optimal closure (\ref{closed-reduced}) is thus completed once  the value function $v(a,b)$ 
is determined, at least to some suitable approximation.   
This calculation is the content of the next section.    

\section{Derivation of the closed reduced equations} 

\subsection{Linear response approximations}     

Solving the Hamilton-Jacobi equation (\ref{HJ}) is difficult in general, especially for a  
closure potential $w$ as complicated as that defined by (\ref{QRsq},\ref{w}).    
We therefore resort to a 
perturbation analysis to obtain approximate solutions $v(a,b)$ in a neighborhood of
statistical equilibrium.  Before presenting a full perturbation calculation, it is  helpful 
to consider two simpler cases in which only quadratic terms
in the cost function, $\L$, or equivalently the Hamiltonian, $\H$, are retained;  this approximation
leads to dissipative linear equations for the resolved variables,  which are
valid nearby statistical equilibrium, as in linear irreversible thermodynamics \cite{dGM}.     
We first consider the case when $b_k = \beta_k$ for all $k \in K_n$, and derive the
linear response of the mean amplitudes, $a_k$ $(k \in K_m)$;    this case represents a further reduction of
the general formulation, in which the variances of the low modes are frozen to their equilibrium values.   
We then consider the alternate
case when $a_k = 0$  for all $k \in K_n$, and derive the linear equations for the relaxation
of the normalized perturbations, $y_k \doteq ( b_k - \beta_k )/\beta_k $ $\; (k \in K_m)$, of the inverse variances;
this case represents homogeneous turbulence near equilibrium, in which there is mean vorticity
field vanishes.   

Setting $b_k = \beta_k$ for all $k \in K_n$ and retaining only quadratic order terms yields a cost function  
\[
\L (a, \dot{a} ) \, = \, \sum_{ k  \in K_m^+} \beta_k |\dot{a}_k|^2 \, + \, \gamma_k |a_k|^2   \, , 
\]  
where   
\begin{equation}  \label{gamma} 
\gamma_k  \doteq  2 \sum_{ p+q=k}  \mu_{p,q}^2  \, c(p,q)^2 \frac{\beta_k^2}{\beta_p \beta_q} \, ;
\end{equation} 
this formula for the dimensionless coefficients $\gamma_k$ follows from (\ref{c-beta-identity}).        
The associated Hamilton-Jacobi equation (\ref{HJ})  for $v=v(a)$ is then 
\[
\sum_k \beta_k^{-1} \left|  \frac{ \d v}{\d a_k^*}   \right|^2  \, - \, \gamma_k |a_k|^2  \, = \, 0 \,  ,
\hspace{1cm}  \mbox{ with } \;\; v(0) = \frac{\d v}{\d a^*} (0) = 0 \, .  
\]
Since this Hamilton-Jacobi equation is both separable and quadratic,  its solution is a quadratic form
\[
v(a) \, = \,  \sum_k M_k  |a_k|^2   \, .  
\]
It follows immediately that $M_k = \sqrt{\gamma_k/\beta_k}$.   In this case the conjugate variables
are  $\phi_k = \beta_k \dot{a}_k$, and hence the near-equilibrium linear relaxation equations for the 
mean amplitudes are simply
\begin{equation}   \label{linear-relax-a}  
\frac{ d a_k}{dt}   \, = \, -  \sqrt{\frac{\gamma_k}{\beta_k} } \, a_k    \hspace{1.5cm} ( \, k \in K_m^+ \, ) .    
\end{equation} 

The key consequence of the decoupled system (\ref{linear-relax-a}) is that the linear dissipative term in the  reduced
equations for the coarse-grained, mean vorticity, $\bar{\zeta}(x,t) = \sum_k a_k(t)  e^{ik \cdot x} $, 
is given by a pseudo-differential operator with symbol
\begin{equation}  \label{sigma}
\sigma_k = \sqrt{\frac{\gamma_k}{\beta_k} }  \hspace{5mm}  \mbox{ where }  \;\;\;\; 
   \gamma_k \approx  \;  C_1 |k|^2 \log C_2 |k| + C_3 \, , 
\end{equation} 
for  certain dimensionless positive constants  $C_1,C_2,C_3$.   
The approximate $k$-dependence displayed here is demonstrated analytically in Appendix A under the restriction
that the sequences of primitive weights  is  constant, $\mu_{p,q}=\mu$,  and $\beta_k = \beta$, that is,  $\theta=0$.   
Numerical evaluation of the sum (\ref{gamma}) shows that  this behavior holds quite generally under mild conditions on the
weight sequence, and also for nonzero $\theta$.  
The notable feature of this dissipation operator is that it is not a standard diffusion,
since  $\sigma_k $ scales with $|k| \sqrt{\log |k|}$ for large $|k|$, rather than with $|k|^2$.     

Alternatively, setting $a_k = 0$ for all $k \in K_n$, and retaining only the quadratic terms in the
normalized perturbations, $y_k = (b_k - \beta_k) / \beta_k $, of inverse variances  yields the cost function 
\[
\L (y,\dot{y})   \, = \, \sum_{k \in K_m^+}  \frac{\dot{y}_k^2}{2 } \, + \,   
     \frac{1}{3} \sum_{k \in K_n} \sum_{p+q=k} \nu^2_{p,q} 
            \frac{ [ \, c(q,k) \beta_p y_p + c(k,p) \beta_q y_q - c(p,q) \beta_k y_k \, ]^2 } {\beta_p \beta_q \beta_k }  \, . 
\] 
The closure potential in this cost function is expressible as a quadratic form in $y \in K_m^+$, namely, 
\[
\L (y,\dot{y})   \, = \, \sum_{k \in K_m^+}  \frac{\dot{y}_k^2}{2 } \, + \,   \frac{1}{2} \sum_{j,k \in K_m^+} D_{j,k} \, y_j y_k  \, .  
\]  
introducing the matrix of coefficients, $  D_{j,k} \, =  \, (\gamma'_k/\beta_k) \, \delta_{j,k} \, + \, G_{j,k}   \, $, 
where    
\begin{equation}   \label{gamma-prime} 
\gamma'_k  \doteq  4 \sum_{ p+q=k}  \nu_{p,q}^2  \, c(p,q)^2 \frac{\beta_k^2}{\beta_p \beta_q} \, ,
\end{equation} 
and  
\[
G_{j,k} \doteq  4 \nu^2_{j,k}  \left\{ \frac{  c(j,j+k) c(k,j+k) }{ \beta_{j+k} }    + 
    \frac{c(j,j-k) c(k,j-k) }{ \beta_{j-k} }  \right\}    \;\;\;\;   \mbox { for } j \neq k  \, ,  
\]
and $G_{k,k} \doteq 4 \nu^2_{k,k} c(k,2k)^2 / \beta_{2k} $ .
The desired solution of the Hamilton-Jacobi equation for this case is a quadratic form,
\(
v(b) \, = \, \frac{1}{2} \sum_{j,k}  R_{j,k} \, y_j y_k   \, , 
\)
whose matrix $R=( R_{j,k})$ satisfies the matrix Riccati equation, $ R^2 = D$.  
  Since $D$ is a symmetric, positive-definite matrix, its has a unique
symmetric, positive definite square root, $R = \sqrt{D}$.   The relaxation of the inverse variances,  
$b_k = \beta_k (1 + y_k)$, to equilibrium for this case of optimal closure is therefore governed by the linear system
\begin{equation}  \label{linear-relax-y} 
\frac{ d y_k}{dt}  \, = \, - \sum_j R_{k,j} y_j   \,   .    
\end{equation}  

The essential content of (\ref{linear-relax-y}) is readily grasped once we notice that the
diagonal elements in the matrix $D$ are large in comparison with the elements in the matrix $G$.   
Indeed, each $\gamma'_k/\beta_k$ given by (\ref{gamma-prime}) is the sum of  $O(n)$ positive 
terms, while each element $G_{j,k}$ is either one term for $j=k$, or a sum of 2 terms for $j \neq k $,
having similar structure and scaling properties.   
Since $D$ is a $d_m \times d_m$ matrix we therefore expect that it will be diagonally dominant when $ n \gg m$.   
When we neglect $G_{j,k} $ in favor of the diagonal, $D_{k,k}$, we get simply
\[
R_{k,k} \approx \sigma'_k \doteq \sqrt{\gamma'_k/\beta_k} \, , 
   \;\;\;\;\;   \mbox{ and }  \;\;\;\;\;  R_{j,k} \approx 0  \;\; \mbox{ for }  \; j \neq k \, .   
\] 
These approximate relaxation rates, $\sigma'_k$, of the inverse variances to equilibrium are entirely analogous 
to those obtained for the mean amplitudes in (\ref{linear-relax-a}), and hence their scaling 
with $|k|$ is also represented by (\ref{sigma}), with some other dimensionless  constants.

\subsection{Perturbation analysis}
      
In this subsection we approximate the solution of the Hamilton-Jacobi equation up to
cubic order terms in a regular
perturbation expansion around statistical equilibrium.     
The partial derivatives of the solution $v=v(a,b)$ that define the irreversible terms in
the closed reduced equations (\ref{closed-reduced}) are thereby expanded up to quadratic
terms.  In this way the dominant nonlinearity in the optimal closure is revealed at the same order
as the quadratic nonlinearity in the Euler equation itself.     
This Taylor expansion is valid under the normalized conditions
\[
 \sqrt{\beta_k} \, | a_k |  \, \ll \, 1  \, ,  \;\;\;\; \frac{| b_k - \beta_k | } {\beta_k} \, \ll \, 1   \, .  
\]

The complicated dependence of the closure potential, $w=w(a,b)$,  in (\ref{w},\ref{QRsq}), on $b$
makes the derivation of the nonlinear reduced equations cumbersome.   For this reason we invoke 
a simplification which rests on the insight gained in the preceding subsection that the diagonal terms in 
the linear relaxation system (\ref{linear-relax-y}) are dominant.     
 To define this approximation, we write the closure potential ({\ref{w}}) as 
\[ 
w(a,b) \, = \,   \Sigma_I + \Sigma_{II} + \Sigma_{III}   \, , 
\] 
introducing a shorthand for the three sums appearing in (\ref{QRsq}), corresponding
to the three independent sets of primitive weights $\lambda_k, \; \mu_{p,q}, \; \nu_{p,q}$, respectively.   
Since $\Sigma_I$ is homogeneous to degree four in $a$, it contributes quartic
terms in $a$ to the Hamilton-Jacobi equation (\ref{HJ}), which are one order higher
than the retained order of our  perturbation expansion;  $\Sigma_I$ is  therefore dropped from the closure potential.   
In $\Sigma_{II}$ and $\Sigma_{III}$ we 
observe that when both $p \in K_n \backslash K_m$ and $q \in K_n \backslash K_m$,
 then $ b_p = \beta_p$ and $ b_q = \beta_q$, and hence the complicated terms in those sums 
 greatly simplify thanks to the relation (\ref{c-beta-identity}).          
 For $n \gg m$, this case occurs for the majority of the indices $p,q$;  indeed,  for each $k$
 the  relative proportion of the index pairs, $(p,q)$, with $p+q=k$, for which it does not hold is 
 bounded by $C  m/n \ll 1$.    On this basis  we set $b_p = \beta_p, b_q= \beta_q$ in $\Sigma_{II}$ and $\Sigma_{III}$.
 Under this approximation the inverse variance for each resolved mode $k$ effectively interacts with an equilibrium ``bath" 
 of other modes via triads, $p+q=k$, and hence  the effective closure potential becomes separable over $k$.  Specifically, 
 \begin{equation}   \label{w-approx}
 w(a,b) \, \approx \,    \sum_{k \in K_m^+}  \;  \, \gamma_k \,  |a_k|^2 \,     
   \, + \, \gamma'_k \,    \frac{( b_k - \beta_k )^2}{ 2 \beta_k^2 \, b_k}  \, , 
 \end{equation}   
 where $\gamma_k $ and $\gamma'_k $ are given by (\ref{gamma}) and (\ref{gamma-prime}), respectively.

In Appendix B the Hamilton-Jacobi equation (\ref{HJ}) with closure potential (\ref{w-approx}) is solved up to cubic order in the
mean amplitudes, $a_k$, and the inverse variance perturbations, $y_k = (b_k - \beta_k)/\beta_k$, and the
governing equations for the optimal closure (\ref{closed-reduced}) are thereby calculated to this order of approximation.  
The result is as follows.   In terms of the derived constants 
\[
\sigma_k  \doteq \sqrt{ \gamma_k / \beta_k }  \, ,  \hspace{1cm}   
\sigma'_k  \doteq \sqrt{ \gamma'_k / \beta_k }  \, ,   \hspace{1cm}  
 \chi_k \doteq \frac{\sqrt{\gamma_k} } { 2\sqrt{\gamma_k} + \sqrt{\gamma'_k } }  \, .   
\]
and the modifying interaction coefficients 
\[
d(p,q) \doteq \frac{ c(q,k) \beta_p \sigma_p + c(k,p) \beta_q \sigma_q - c(p,q) \beta_k \sigma_k}
                               { ( \sigma_p + \sigma_q  + \sigma_k ) \beta_k  }   \, . 
\]
 the governing equations for the approximated optimal closure are  
\begin{equation}     \label{c-r-a}  
\frac{d a_k}{dt} +  \sum_{p+q=k} \left[ \,  c(p,q) +  d(p,q) \,  \right] a_p a_q \, = \,  
       - \sigma_k \left[ \, 1  -  ( 1 - \chi_k ) y_k \,  \right] \,   a_k 
\end{equation} 
\begin{equation}      \label{c-r-y} 
\frac{d y_k}{dt}  \,  =  \,   - \,  \sigma_k \chi_k \beta_k  |a_k|^2 \,
       -  \sigma'_k  \left[  \,  1  +   \frac{y_k}{2 } \, \right] y_k    \, . 
\end{equation}  
This closed system of reduced equations for the means and inverse variances of the resolved low modes 
is the main result of the paper.      
 
 \subsection{Some properties of the closure}
 
The dominant dissipative mechanisms in the reduced equations (\ref{c-r-a},\ref{c-r-y}) are the linear terms
for the mean and inverse variance perturbation of mode $k$, which are scaled by $\sigma_k$ and $\sigma'_k$, respectively. 
In order to clarify the dissipative structure of the closure   (\ref{c-r-a},\ref{c-r-y}) it is
worthwhile to calculate the entropy production of the macroscopic dynamics.   The
entropy relativized to equilibrium is given by 
\begin{eqnarray}  \label{entropy}
s(a,b) =   - \left\la \, \log \frac{\rhotilde (a,b)}{\rho_{eq}} \, |  \, \rhotilde(a,b) \, \right\ra  
    &=&  - \sum_{k \in K_m^+} \beta_k |a_k|^2 + \log \frac{b_k}{\beta_k} -1 + \frac{\beta_k}{b_k}   \nonumber \\
  & \approx & - \sum_{k \in K_m^+} \beta_k |a_k|^2 + \frac{(b_k - \beta_k)^2}{2 \beta_k^2}  \, ,  
\end{eqnarray}
where the final  expression is an expansion in $b-\beta$ that neglects terms  $ O ( |b - \beta|^3 ) $.   
Substituting the closed reduced equations (\ref{c-r-a},\ref{c-r-y}) into the time derivative of this
entropy yields 
\begin{equation}  \label{ent-prod}  
\frac{d s}{dt}  \, = \,  -  \sum_{k \in K_m^+}  \beta_k a^*_k \frac{d a_k}{dt} \, + \, 
               \frac{(b_k - \beta_k)}{ 2 b_k^2} \frac{d b_k}{dt}   
    \;  \approx \;   \sum_{k \in K_m^+} \sigma_k \beta_k  |a_k|^2   +  \sigma'_k  \frac{(b_k - \beta_k)^2}{ 2 \beta_k^2} \, , 
\end{equation} 
retaining only the leading order terms in this expression for entropy production.   Thus, we verify
that entropy is monotonically increasing in forward time, within the range of validity of the perturbation
analysis.    The entropy production formula also reveals the relative contributions from the nonequilibrium
mean, via $|a_k|^2$, and nonequilibrium variance, via $(b_k -\beta_k)^2$, of the various
modes $k$.   

The closure (\ref{c-r-a},\ref{c-r-y}) contains two nonlinearities in its irreversible terms, beyond
the dependence of the dissipation rates on the nonequilibrium variance perturbation.    Namely,  there are
modified triad interactions between the mean amplitudes regulated by $d(p,q)$, and couplings
between the mean and variances of each mode $k$ scaled by $\chi_k$.  

The coupling between the variance and mean captured by the first term of (\ref{c-r-y}) is a noteworthy
prediction of the optimal closure.      Recalling that $b_k$ is
the inverse of variance in the statistical model, the negative term in (\ref{c-r-y}) that
is proportional to $|a_k|^2$ represents a positive source of variance in the presence of a nonequilibrium
(nonzero) mean $a_k$, and this term can counteract the relaxation of the variance disturbance.    
It is sometimes observed in simulations, such as those 
documented in Section 6, that the variance of a mode with substantial mean perturbation 
can overshoot its equilibrium value during relaxation.    This effect is revealed when we 
 ignore the modal interactions  in (\ref{c-r-a},\ref{c-r-y}) and form  
 the $ 2 \times 2$ system of first-order ordinary differential equations 
in the variables $|a_k|^2$ and $y_k$ for each particular mode $k$.  
In this phase plane some orbits with initially large enough $\beta_k |a_k|^2$ and positive
$y_k$ cross $y_k=0$ before relaxing to 0; that is, a mode with relatively large mean and
small variance may exceed equilibrium variance in the process of its relaxation to equilibrium.

The modified interactions between the mean resolved modes may be put into a more
illuminating form as follows.   We calculate   
\begin{eqnarray*}  
c(p,q) +  d(p,q) &=& \frac{  \sigma_p  [ \, c(p,q)  \beta_k  +  c(q,k) \beta_p \, ]  + 
                                  \sigma_q [ \,  c(p,q) \beta_k + c(k,p) \beta_q \, ] }
                              {     (  \sigma_p + \sigma_q + \sigma_k )   \beta_k }       \, ,                         
\end{eqnarray*}   
and we observe symmetry between $p$ and $q$ in the terms of this expression.
On this basis we introduce  asymmetric interaction coefficients
\begin{equation} \label{coeff-asym}
\bar{c}(p,q)   \doteq   \frac{ 2 \sigma_p} { \sigma_p + \sigma_q + \sigma_k } 
                       \left[ \, c(p,q) + c(q,k) \frac{\beta_p}{\beta_k} \,    \right]      \hspace{2cm}  
                             ( \mbox{ for } \, p+q=k \, ) \, 
\end{equation}  
and we rewrite the interaction terms in the reduced equations (\ref{c-r-a})  in the simpler form:
\[
\sum_{p+q=k} \left[ \,  c(p,q) +  d(p,q) \,  \right] a_p a_q \, =   \sum_{p+q=k} \,  \bar{c}(p,q)  \,  a_p a_q \, 
\]
A further calculation exhibits the relationship of $   \bar{c}(p,q)$ to the primitive interactions in the vorticity dynamics
(\ref{microdynamics}) itself;  namely,
\[
\bar{c}(p,q) = - \frac{ p \times q}{|q|^2} \cdot \frac{ \sigma_p} { \sigma_p + \sigma_q + \sigma_k } 
    \left[  \,  1 - \frac{|q|^2} {|p|^2}  \, + \, \frac{\beta_p}{\beta_k}  \left( 1 - \frac{|q|^2} {|k|^2} \right) \, \right]  \, . 
\]
When $|q| \ll |p|$ and hence  these
interaction terms are relatively large, the factor multiplying the primitive interaction coefficient is nearly unity,
since in that case $|q| \ll |k|, \; |p| \sim |k|, \; \beta_p \sim \beta_k, \; \sigma_q \ll \sigma_p \sim \sigma_k$;  
for these triads of wavevectors,  $\bar{c}(p,q) \approx c(p,q)$, suggesting that the modification of the
interactions is weak.    
Recalling that $p$ is the wavevector associated with the advected mean vorticity field and $q$ is the
wavevector associated with the mean streamfunction, this heuristic result suggests that the
mean vorticity is strained by the mean flow in much the same way as vorticity is strained by its induced
flow in the deterministic dynamics itself.     This rough  argument notwithstanding, 
the optimal closure theory makes a definite prediction about how averaging over the
unresolved fluctuations in vorticity leads to modified interactions between the modes of the mean vorticity
field.

\section{Comparison with direct numerical simulations} 

In this section we present the results of some 
numerical experiments designed to test the predictive skill of the optimal coarse-grained dynamics.  
These simulations are carried out on a modestly truncated system, in which $n=32$ and $m=4$.    Thus, a 
microdynamics with $d_n=2112$ degrees of freedom is reduced to a macrodynamics in $d_m=40$ 
resolved variables.    Such a truncation may appear to be severe from the point of view of computational fluid dynamics.  
Nonetheless, it serves as a valid platform for testing the optimal closure, since this model reduction
technique is applicable to the statistical evolution to any (isolated) Hamiltonian dynamical system of high dimension \cite{Turk}.   
Furthermore, the requirement that the ensembles of fully-resolved solutions should be statistically converged requires that
the number of realizations, $N$,  in the ensembles be very large, which puts a practical constraint on $n$.

For the direct numerical simulations we employ the dealiased Fourier pseudospectral method with
$96^2$ collocation points.    The fourth-order Runge-Kutta method is used with fixed time step, $\Delta t$ = 0.002, 
chosen to ensure conservation of energy and enstrophy.   
The code is parallelized with OpenMP in order to handle the large ensemble size.    
The optimal closure itself is computed by the Runge-Kutta method applied to the low-dimensional system 
of closed reduced equations (\ref{c-r-a},\ref{c-r-y}).  

\subsection{Initializing the ensembles} 
To generate an initial ensemble, $\rho^0 = \rhotilde(z;a^0,b^0)$,
 with specified (complex) means $a_k^0$ and variances $1/b_k^0$
for the resolved modes $k \in K_m^+$, together with equilibrium means and variances
for the unresolved modes, we proceed as follows.    The energy and enstrophy 
contained in the resolved modes are
\begin{eqnarray}   \label{energy-enstrophy-resolved}
E_m^0 = \frac{1}{2} \sum_{ k \in K_m} |k|^{-2} \left( |a_k^0|^2 + \frac{1}{b_k^0} \right) \, , \hspace{1cm}
Q_m^0 = \frac{1}{2} \sum_{ k \in K_m}   |a_k^0|^2 + \frac{1}{b_k^0}  \, .
\end{eqnarray}  
Since the energy content and enstrophy content of each unresolved mode $k \in K_n \backslash K_m$ 
are identical in the equilibrium state, $\rho_{eq}$, and the initial state, $\rho^0$, 
the specified initial state is guaranteed to be dynamically compatible under statistical relaxation 
with the equilibrium state provided that 
\begin{equation}    \label{beta-theta-eqns}
E_m^0 = \frac{1}{2} \sum_{ k \in  K_m} \frac{1}{\beta |k|^2 + \theta }  \, , \hspace{1cm}
Q_m^0  =  \frac{1}{2} \sum_{ k \in  K_m} \frac{1}{\beta + \theta |k|^{-2} }  \, .   
\end{equation}
These two equations determine the equilibrium parameters, $\beta$ and $\theta$, subject to  
$\beta >0$ and $\theta > -\beta$.   A unique pair $(\beta,\theta)$ exists by virtue of the fact that
these equations are the critical point conditions for the strictly convex minimization 
\[
\min_{\beta,\theta} \; E_m^0 \theta \,+\, Q_m^0 \beta \, - \frac{1}{2} \sum_{ k \in  K_m} \log (\beta + \theta |k|^{-2} )  \, . 
\] 
Thus, when constructing a nonequilibrium initial state for a statistical relaxation experiment,  
the means and variances of the resolved (low) modes may be chosen arbitrarily, 
within the near-equilibrium limitations of the perturbation analysis of Section 5,  and 
from them the compatible equilibrium parameters $\beta$ and $\theta$ may be calculated.

\subsection{Testing the Gaussian trial densities}

Before evaluating the error between fully-resolved ensembles and the optimal closed reduced dynamics, 
we first check the normality assumption of the model.   To do so we calculate the correlation coefficients
of any two modes, as well as the skewness and excess kurtosis of each individual mode, in light of the basic
assumption of our model reduction that all of these quantities vanish with respect to the  trial densities
(\ref{trial-densities}).   
An initial ensemble is generated by random sampling a quasi-equilibrium density with $\beta=1$ and $\theta=0$, and having
perturbations in mean and variance, namely,  $|a_k^0| $ and $|b_k^0 -\beta|$, that vary from $0.07$ to $0.98$ and from
$0.10$ to $0.14$, respectively;  the ensemble contains 40,000 realizations.  
For the fully-resolved nonequilibrium ensemble with this initial condition,  
the correlations between resolved modes is found to remain very small throughout its relaxation to equilibrium.    
For example, the four correlations between the real and imaginary parts of modes $(-4,0)$ and $(-3,1)$ all lie 
in the interval $[-0.05 , +0.05]$  uniformly over the computed relaxation.   Over the same time interval,
 the skewness of mode  $(-4,0)$ never exceeds $0.03$ in magnitude, and the excess kurtosis of mode  $(-4,0)$ 
 never exceeds $0.06$.     
These ranges are typical for the other resolved modes.   Similarly, the departures from normality of the 
unresolved modes, which the model sets to equilibrium statistics, are found to be small under the spectrally-truncated 
microdynamics (\ref{microdynamics}).    
These tests support the choice of the trial densities (\ref{trial-densities}) to approximate
the true statistics of the evolving nonequilibrium statistical state.

\subsection{Quantifying the error between model and DNS}   

To quantify the goodness-of-fit between a reduced model trajectory  and the corresponding ensemble
of fully-resolved solutions, a natural criterion is supplied by the relative entropy.  
Namely, the Kullback-Leibler divergence \cite{Kullback,CT} between the evolving quasi-equilibrium density, $\rhotilde(a(t), b(t))$,
 and the exactly propagated density, $\rho^{dns}(t)$, is 
 \begin{eqnarray}   \label{k-l} 
 D( \rho^{dns} \, |  \, \rhotilde ) & \doteq &  
                    \left\la   \log \frac{\rho^{dns}}{\rhotilde} \,  |  \,  \rho^{dns}   \right\ra   \nonumber \\
     &=&  \sum_{k \in K_n^+}   b_k | a_k^{dns} - a_k |^2 \, + \, 
     \log \frac{ b_k^{dns}}{b_k} + \frac{b_k}{ b_k^{dns}} -1     \nonumber \\
     & \approx &  \sum_{k \in K_n^+}   b_k | a_k^{dns} - a_k |^2 \, + \, 
      \frac{ (b_k^{dns} - b_k)^2}{ 2 ( b_k^{dns})^2}   \, ,  
 \end{eqnarray}  
 where the approximation holds when $b_k \approx b_k^{dns}$;  $a_k^{dns}$ and $b_k^{dns}$
 denote the means and variances derived from $\rho^{dns}$.     Motivated by this calculation,
 we use the following nondimensional squared norm to assess the fit between model predictions and direct numerical
 simulations of ensembles:  
 \begin{equation}    \label{squared-norm}  
 \Delta_m (  \rho^{dns} \, |  \, \rhotilde ) \,  \doteq \,     
 \frac{1}{d_m}  \sum_{k \in K_m^+}   \beta_k | a_k^{dns} - a_k |^2 \, + \, 
      \frac{ (b_k^{dns} - b_k)^2}{ 2 \beta_k^2}  \, . 
 \end{equation}  
 This squared error per mode retains the intrinsic weighting between mean and variance deviations dictated by
 the relative entropy formula (\ref{k-l}), but for simplicity it replaces the nonequilibrium inverse variances that scale the 
 squared differences by their equilibrium values.  It pertains to the  $d_m = 2m (m+1)$  
 resolved modes and does not include any errors in the unresolved modes;  as such it measures  the
 skill of the optimal closure to predict the evolution of the coarse-grained vorticity field.  
 The mean squared error averaged over some time interval, $0 \le t \le T$, is denoted by 
 $\bar{\Delta}_m(\rho^{dns} | \rhotilde) $.  This time-averaged, relative-entropy based, squared norm is our principal metric
 for quantifying the goodness-of-fit of the optimal closure.

Table~\ref{tab:pos}  gives the parameters for a set of 5 initial conditions having positive inverse temperatures, $\theta$;
across the columns $\beta$ decreases from $0.99$ to $0.76$, while $\theta$ increases from $0.013$ to $0.61$.   
Table~\ref{tab:neg} shows the parameters for a similar set of  5 initial conditions having negative inverse temperatures, $\theta$;
across the columns $\beta$ decreases from $1.21$ to $0.61$, while $\theta$ decreases from $-0.025$ to $-0.15$.      
In each of these tables the magnitude of the initial perturbation  increases across the columns;
these magnitudes are quantified by 
\[
\Delta_m( \rho^0 \, | \, \rho_{eq} )  \doteq  
    \frac{1}{d_m} \sum_{k \in K_m^+}  |a_k^0|^2 \beta_k +  \frac{ |b_k^0 - \beta_k|^2 }{2 \beta_k^2} \, ,
\] 
which, as in (\ref{entropy}),  is the approximated relative entropy per mode between
the initial density and the equilibrium density.  
These tables also list the maximum relativized perturbations in the mean and inverse variances 
over the disturbed modes $k \in K_m^+$.

\begin{table}[htbp]
\caption{Parameters used for five positive temperature ensembles with increasing perturbation magnitudes.    
The initial perturbation is quantified by the relative-entropy-based squared norm $ \Delta_m(\rho^0 \, |  \, \rho_{eq}  \, ) $.    }
\label{tab:pos}
\begin{center}
\begin{tabular}{|c|l|l|l|l|l|} \hline
Total Energy 	&  28 	& 28	& 29	& 30	& 32 \\ \hline
Total Enstrophy 	&  $9.5 \times 10^3$	& $9.6 \times 10^3$	& $1.0 \times 10^4$	& $ 1.1\times 10^4$	&  $1.2 \times 10^4$\\ \hline
$\max{|a_k^0| \sqrt{\beta_k} }$ 	&  0.30	& 0.50	& 0.98	&1.5	&1.8 \\ \hline	
$\max |b_k^0 - \beta_k|/\beta_k$ &  0.02	&0.06&	0.26	&0.92	&2.3 \\ \hline 
 $\Delta_m( \rho^0 \, | \, \rho_{eq} ) $ 
 &  0.018	&0.050	&0.216	&0.803	&2.78	\\  \hline 
\end{tabular}
\end{center}  
\end{table}

\begin{table}[htbp]
\caption{Parameters used for five negative temperature ensembles with increasing perturbation magnitudes.   
 The initial perturbation is quantified by the relative-entropy-based squared norm  $ \Delta_m( \rho^0 \, |  \, \rho_{eq}  \, ) $.    }
\label{tab:neg}
\begin{center}
\begin{tabular}{|c|l|l|l|l|l|} \hline
Total Energy 	&  23 	& 23	& 24	& 34	& 44 \\ \hline
Total Enstrophy 	&  $7.7 \times 10^3$	& $7.5 \times 10^3$	& $8.1 \times 10^3$	& $ 1.1\times 10^4$	&  $1.4 \times 10^4$\\ \hline
$\max{|a_k^0| \sqrt{\beta_k} }$ & 0.38 &	0.57	&0.92	& 1.3 &	1.4 \\ \hline	
$\max |b_k^0 - \beta_k|/\beta_k$ &  0.08 &	0.20	& 0.66	&2.1	& 3.7	  \\ \hline 
 $\Delta_m( \rho^0 \, | \, \rho_{eq} ) $  & 0.058	 &0.140	 &0.454	& 1.87	& 4.68	\\ 
\hline   
\end{tabular}
\end{center}  
\end{table}

\begin{figure}
  \centering
        \includegraphics[width=0.49\textwidth]{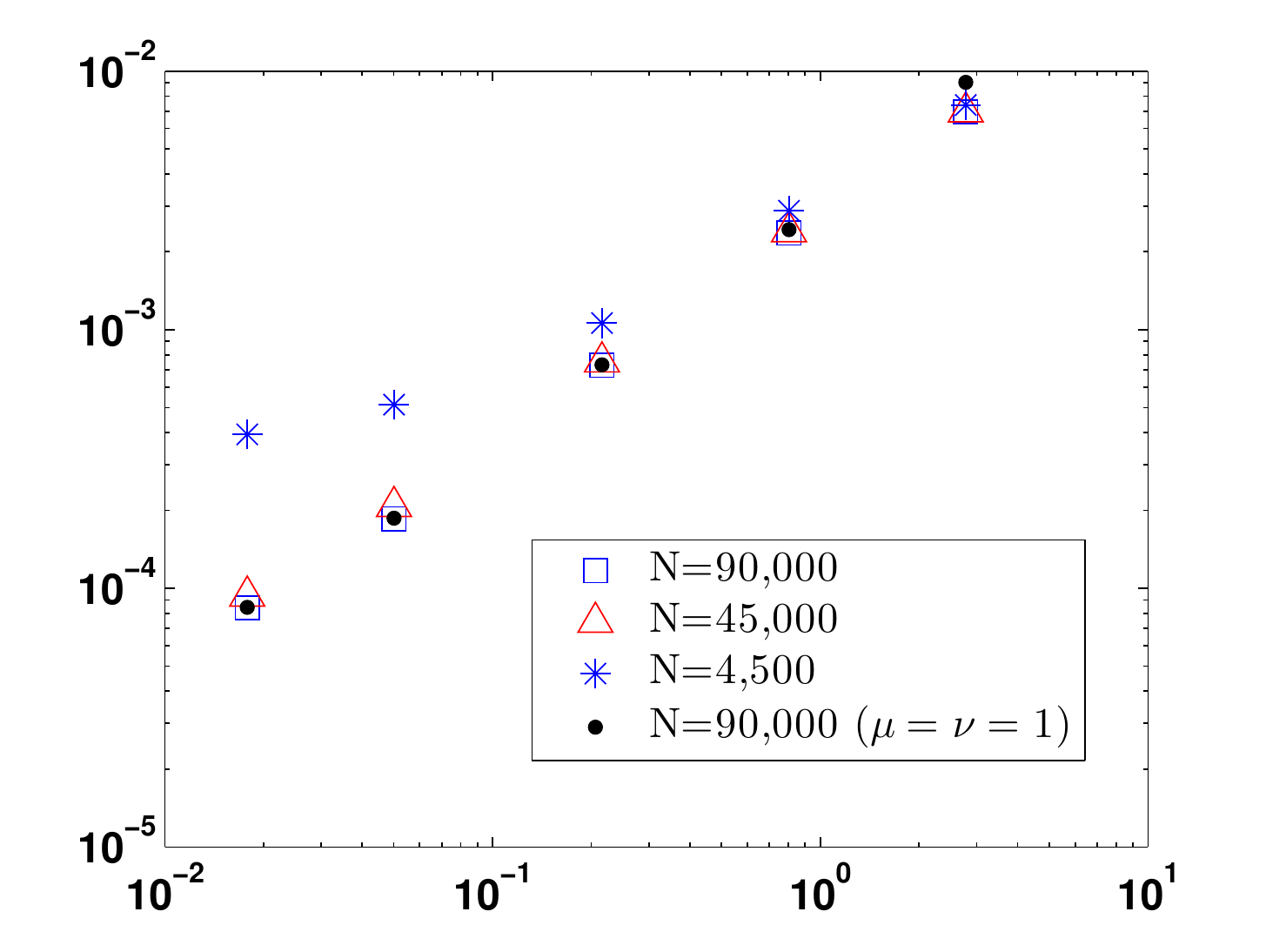}
       \includegraphics[width=0.49\textwidth]{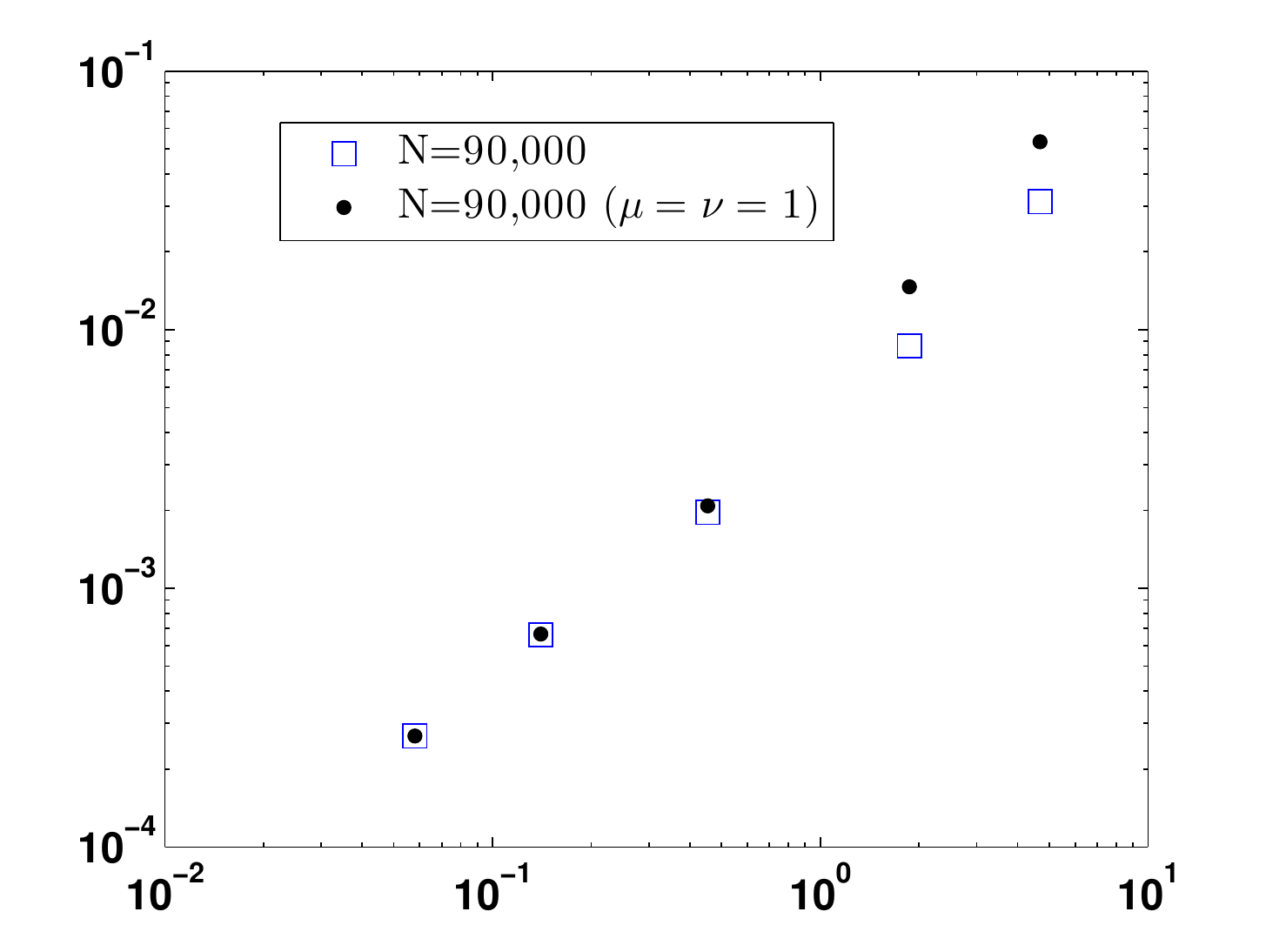}
     \caption{Squared norm $\bar{\Delta}_m( \rho^{dns}  \, | \, \tilde{\rho} ) $
     versus perturbation magnitude $\Delta_m( \rho^0 \, | \, \rho_{eq} ) $. 
		 Left: Positive temperature as in Table 1; Right:   Negative temperature as in Table 2. }
    \label{f:error-perb}
    \end{figure}

In Figure \ref{f:error-perb}  the time-averaged  squared error,    $ \bar{\Delta}_m(\rho^{dns} \, |  \, \rhotilde \, ) $, 
is plotted against the squared norm of the initial perturbation, 
$\Delta_m( \rho^0 \, | \, \rho_{eq} ) $.   For all the cases displayed in Figure \ref{f:error-perb} 
the time averaging is taken over the interval, $0 \le t \le 3$, since the slowest modes almost completely relax by this time. 
The left panel of this figure demonstrates the necessity of taking large ensembles, in that it contrasts the computed error for $N=4500$ 
random realizations against $N=45,000$ and $N=90,000$.    
The larger errors associated with smaller $N$ are attributable to sampling error rather than model error.  
Statistical convergence is obtained at $N=90,000$, in that the resulting error
hardly decreases with further increased sampling.   
We therefore take the results for $N=90,000$ as representative of the true discrepancy between ensemble DNS and the optimal closure model.  

The coarse-grained dynamics (\ref{c-r-a},\ref{c-r-y}) used to propagate the model solution, $\rhotilde(a(t),b(t))$, for
the tests in Figure \ref{f:error-perb} contain two adjustable parameters $\mu$ and $\nu$, which scale $\gamma_k$ 
and $\gamma'_k$, appearing in (\ref{gamma}) and (\ref{gamma-prime}), respectively.   We set $\mu_{p,q} = \mu$
over the wavevector indices $p,q \in K_n $, with $p+q=k$, in the sum defining $\gamma_k$, and similarly $\nu_{p,q} = \nu$
for $\gamma'_k$,  and we seek those parameters, $\mu$ and $\nu$, that minimize 
the time-averaged squared error between the optimal closure model and DNS.    
Numerical evaluations of the sums $\gamma_k$ and  $\gamma'_k$ for various sequences of 
primitive weights $\mu_{p,q}, \nu_{p,q}$ shows that they are relatively insensitive to variations of the primitive weights 
over their indices $p$ and $q$.   It is therefore doubtful that a much better fit between the model and DNS 
could be achieved by considering more complicated choices of the primitive weights.     In the next subsection,
the universal choice, $\mu_{p,q} = \nu_{p,q}= 1$ for all $p$ and $q$,  is discussed, and the results
for  $\mu=\nu=1$ are also plotted in Figure \ref{f:error-perb}.

We find excellent agreement for small initial perturbations.     For instance,  the first data column 
of Table~\ref{tab:pos} gives $\Delta_m(\rho^0 | \rho_{eq}) = 0.018 $, and we may ascribe almost all this squared error to the mean,
since the variance perturbation is very small in this case.  The initial mean perturbation averaged over all disturbed modes is 
thus $13\%$, while the maximum initial perturbation among those disturbed modes is $30\%$.   
On the one hand, when we ascribe all the time-averaged squared-norm error,  
$\bar{\Delta}_m( \rho^{dns}  \, | \, \tilde{\rho} ) = 8.4 \times 10^{-5}$,  to the mean, we obtain a relative error in mean of less than $1\%$.    
On the other other hand, when we ascribe the error to the variance alone, 
we obtain a relative error in variance of less than $2\%$.   Together these small percentages provide upper bounds on the relative error in means and variances averaged over the resolved modes and the time interval of relaxation.       

Good agreement between the large-ensemble DNS and the optimal closure persists for perturbations in mean and variance
of the resolved modes having magnitudes up to order unity with respect to the equilibrium ensemble.      
For instance, in the fourth data column of Table~\ref{tab:pos} the initial perturbation has roughly equal parts in the mean and variance and these correspond to $67\%$ and $83\%$, respectively, relative to equilibrium.  
In this case, when we ascribe equal parts of the squared error,  $\bar{\Delta}_m( \rho^{dns}  \, | \, \tilde{\rho} ) = 2.4 \times 10^{-3}$, 
to mean and variance,  we infer relative errors in mean of $3.4\%$ and in variance of $4.8\%$.    
The fifth data column of  Table~\ref{tab:pos} is included to indicate the limitation of the theory for much larger amplitudes,
 as is anticipated in light of the derivation of the optimal closure via a perturbation expansion in amplitude around
 statistical equilibrium.    In this case the maximum mean and variance perturbations are around twice the normalizing values for equilibrium, and an estimate ascribing equal parts of the squared error to the means and variances yields roughly $6\%$ and $8\%$ 
for the mean and variance errors, respectively.     
 
 Analogous results are obtained for the set of comparisons given in Table~\ref{tab:neg}, where the initial conditions are taken so
 that the inverse temperatures are negative.   As is well known in the statistical equilibrium context, negative temperature states
 occur when energy concentrates in the low modes \cite{KM}.      In an equilibrium theory in which the energy invariant is treated 
  microcanonically, negative temperature states develop large-scale mean flows, identified as coherent structures in the fluctuating 
  vorticity field \cite{MW}.   It is known that the equivalence of ensembles with respect to the energy breaks down in the negative temperature 
  regime, so that the canonical ensemble does not produce stable coherent structures \cite{EHT,Turk-lesHouches,BV}.      
 In the current paper, both the energy and enstrophy invariants are treated canonically to avoid distracting 
 complications in the development of the closure theory.    Consequently, the reference statistical equilibrium states
 have increasingly large variance in the lowest modes as $\theta$ becomes increasingly negative, while the mean of all modes is 
 zero.     The comparisons given in Table~\ref{tab:neg} pertain to relaxation toward states of this kind from initial perturbations in the resolved modes with increasing amplitude, especially in the variance.    For instance, in the third data column the 
 initial perturbation relativized to equilibrium and averaged over the resolved modes is about $57\%$ in mean and $49\%$ in variance.
 If the relative-entropy-based squared error is ascribed equally to mean and variance, the  mean and variance errors are approximately
 $3\%$ and $4\%$, respectively.    From this agreement we conclude that the optimal closure continues to give a good approximation
 to the actual statistical evolution in the negative temperature regime.     The relative errors in the negative temperature regime, 
 however, grow more rapidly with increasing perturbation amplitude than in the positive temperature regime.    For the final two data
 columns in  Table~\ref{tab:neg} the disturbances in mean and, especially, variance are quite large, resulting in substantial
 disagreement.    To what extent these deviations are due to the nature of the negative temperature states as opposed to the limitations of the perturbation analysis is not clear.

\begin{figure}
\centering
        \includegraphics[width=0.49\textwidth]{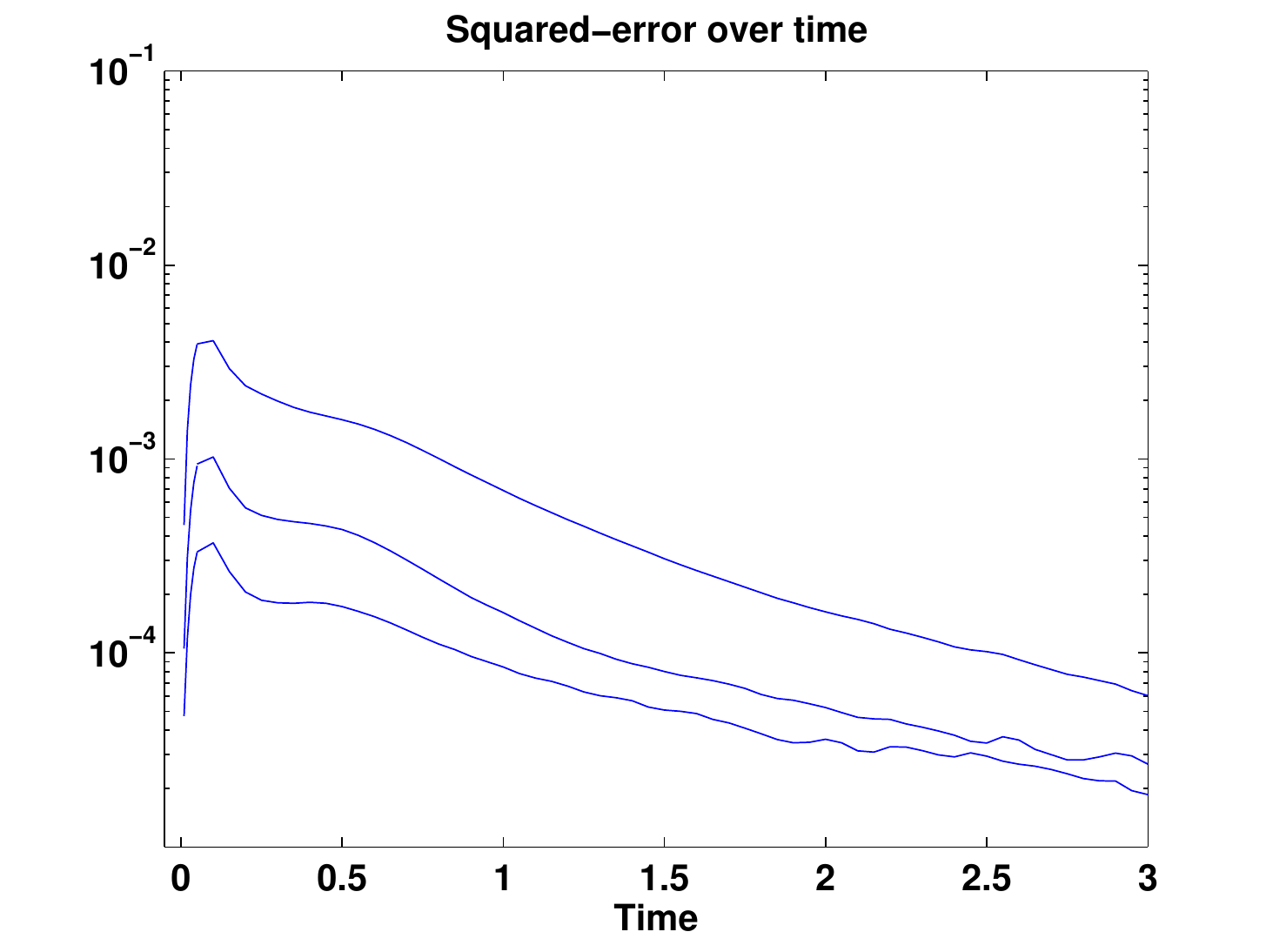}
       \includegraphics[width=0.49\textwidth]{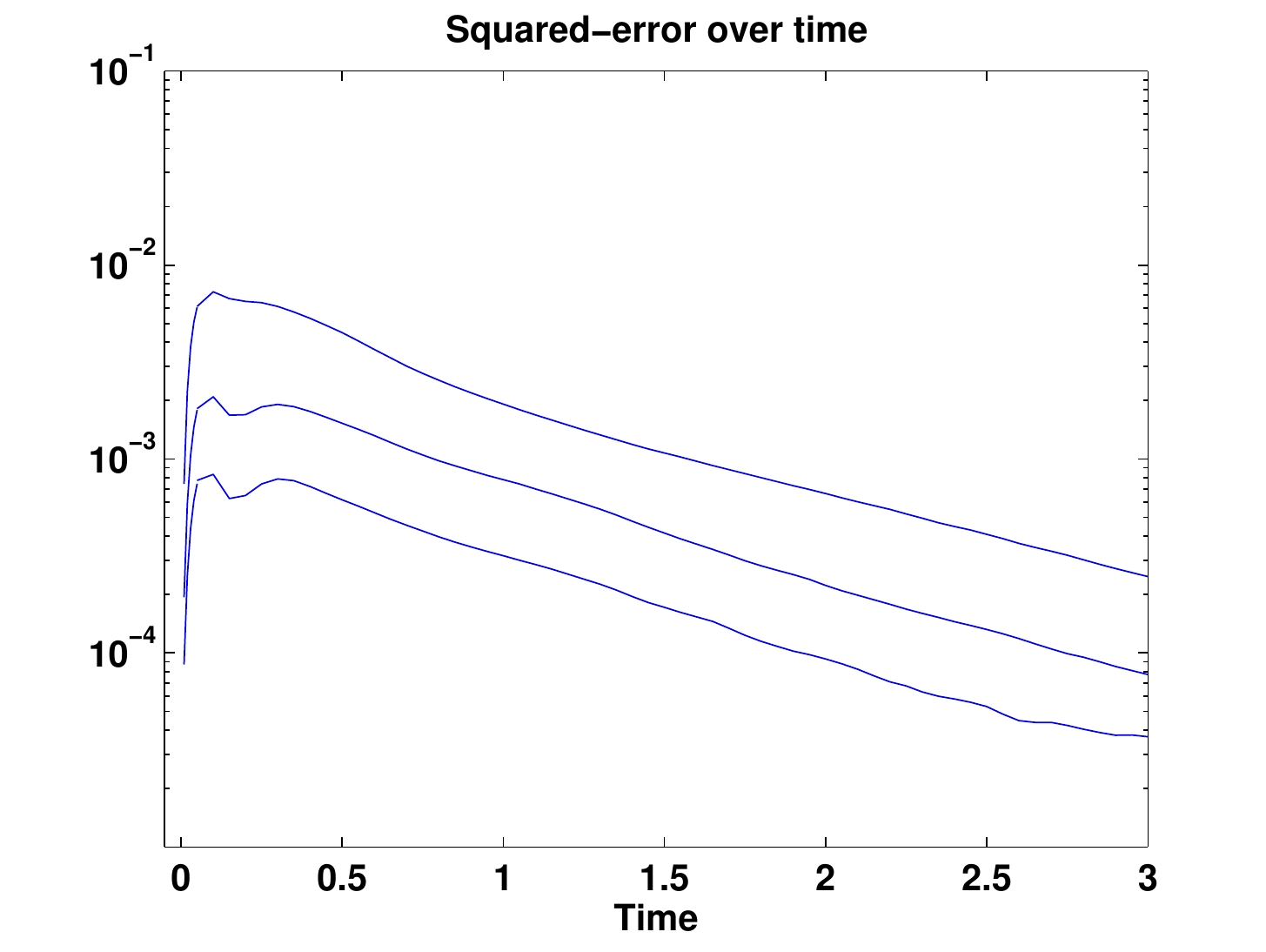}
     \caption{Evolution of the relative-entropy-based squared-error versus time. 
         Each line corresponds to one of the first three columns in Table~\ref{tab:pos} or ~\ref{tab:neg}.
		 Left: Positive temperature; Right:   Negative temperature. }
    \label{f:entropy_time}
    \end{figure}

Setting aside time-averaging in the goodness-of-fit metric,   Figure \ref{f:entropy_time} plots the time evolution of the 
relative-entropy-based squared errors, $\Delta_m(\rho^{dns}(t)  | \rhotilde (t))$, for the three test cases with the smallest
disturbances, in both the positive and negative temperature regimes.       The conspicuous feature of all these plots is
the short period of adjustment from the initial nonequilibrium state during which the squared error grows rapidly and
then saturates before decaying almost exponentially as both the DNS ensemble and the optimal closure equilibriate.
The transient phase during which the instantaneous squared errors are near their peak values is shorter in the positive
temperature cases than in the negative temperature cases.   For all the cases shown in Figure \ref{f:entropy_time}
the squared error peaks at about 4 times its time-averaged value over the interval $0 \le t \le 3$.   Thus, the maximum 
instantaneous root-mean-square errors in either mean or variance are about twice as large as the time-averaged errors 
discussed above.

\subsection{Eliminating the adjustable parameters}

Figure \ref{f:error-perb} displays a striking agreement between the prediction of the optimal closure
with best-fit parameters, $\mu$ and $\nu$, and the optimal closure in which both of these adjustable parameters
are set to unity.    In other words, it is found that the adjustable parameters that determine a  closed reduced dynamics
that is closest to DNS are $\mu, \nu \approx 1$.   This result holds for all the test cases in which the optimal closure agrees 
well with DNS, in both the positive and negative temperature regimes.  We conclude that the 
weights in the lack-of-fit Lagrangian (\ref{cost-fn-def}) may be set equal to unity, or equivalently, that the weight operator
$W$ may be taken to be the identity.   Then the Lagrangian is simply the $L^2(\Gamma_n,\rhotilde)$ squared norm 
of the Liouville residual, without separate treatment of its resolved and unresolved components.     

With  $\mu_{p,q} = \nu_{p,q} = 1$ for all $p,q$, we obtain closed reduced equations (\ref{c-r-a}, \ref{c-r-y})
 that are entirely free of adjustable parameters.    
In this intrinsic closure $\gamma'_k = 2 \gamma_k$, and hence the 
constants appearing in the coarse-grained equations (\ref{c-r-a},\ref{c-r-y}) satisfy
$\sigma'_k = \sqrt{2} \sigma_k$ and $\chi_k = 1/(2 + \sqrt{2})$.    The dissipative structure of
the closure is therefore entirely determined by the sequence of constants $\gamma_k$ 
defined in (\ref{gamma}).    In Appendix A we derive the approximate analytical formula (\ref{universal-gamma}) for $\gamma_k$  
 when  and $\theta=0$.    This formula also provides a good model for $\gamma_k$
 for positive and negative $\theta$, as may be verified numerically.

\section{Conclusions}    

A dynamically optimal coarse-graining of two-dimensional turbulence has been derived 
using the general statistical-mechanical reduction procedure proposed in \cite{Turk}, and its predictions
have been validated against fully-resolved direct numerical simulations of large ensembles.   
Specifically, the evolution of statistical states for a spectral truncation of incompressible, inviscid fluid dynamics 
onto many Fourier modes has been approximated by a Gaussian statistical model that is parameterized by the means and variances of
a set of low modes.    Unlike traditional closure schemes, which typically rely on moment hierarchies and eddy damping strategies,
this reduction procedure is based on an optimization principle, in which  
a cost functional that quantifies the residual due to coarse-graining is minimized.   
We have included arbitrary nondimensional parameters in the cost functional to weight the
unresolved components of the residual relative to the resolved components.   Information-theoretic
considerations \cite{Kleeman} suggest that all such weights can be set to unity, resulting in an optimal closure that
is free of any adjustable parameters.   This conjecture has been tested in Section 6, where it has been found that 
the optimal closure with unit
weights on all  components is essentially indistinguishable from that obtained by best-fitting those
weights against benchmark numerical solutions.  
  One conclusion of this study, therefore, is that the weights can be set to unity from the outset ---
that is, $W=1$ in (\ref{cost-fn-def}) --- 
making the reduction procedure conceptually and technically simpler,  and yielding a system of closed reduced equations 
 that is completely intrinsic.   This universal coarse-grained dynamics pertains to the statistics of spectrally-truncated ideal flow 
in which the unresolved modes 
are taken to be in statistical equilibrium, and the resolved modes are relatively near equilibrium.  Specifically, 
the tests documented in Section 6 show good agreement with DNS for perturbations of the means up to about one equilibrium standard deviation 
and of the variances up to about one equilibrium variance.     
 
While this paper has presented quantitative tests of the optimal closure for the near equilibrium regime, and for a spectrally-truncated
dynamics having a modest resolution, our analytical derivation of the closed reduced equations for the Gaussian closure
is valid for arbitrary high resolution and far from equilibrium.    That is, the reduced equations (\ref{closed-reduced})
give the dynamically optimal evolution of the means and variances of the resolved modes
within the class of Gaussian models (\ref{trial-densities}), for any the resolutions $n$ and $m$
of the fine and coarse scales, respectively, and without restriction on the amplitudes of the resolved variables.  
Moreover, they have a generic thermodynamic structure and associated properties, such as an entropy production inequality  \cite{Ottinger,Turk}.  These attractive features of the optimal Gaussian closure contrast with the properties of standard moment-based closures, which may not be self-consistent or realizable without suitable amendments or adjustments \cite{Orzsag,McComb}.     
The reduced equations for the optimal closure, however, require solving
the Hamilton-Jacobi equation (\ref{HJ}) to determine the dissipation potential $v(a,b)$.    As is recognized in 
the calculus of variations and optimal control theory, Hamilton-Jacobi equations are not often analytically tractable, and
hence one must rely on some approximations (such as the perturbative analysis used in this paper) or numerical solutions of
the optimization problem itself.   Investigations of realistic turbulence phenomena, which are naturally far from equilbrium, 
therefore would require techniques to handle this Hamilton-Jacobi equation beyond those invoked in this paper.

The optimal closure contains dissipative terms 
not anticipated by other approaches to inhomogeneous two-dimensional turbulence \cite{FO}.    In particular, the modification of the nonlinear
interactions between the mean resolved modes is deduced systematically from the optimization principle, as is the interaction between the mean and variance of each individual resolved mode.    These couplings
in the coarse-grained evolution equations  might not be revealed by
generating and truncating a moment hierarchy.   Further investigations are needed to ascertain how 
other approaches to closure may be incorporated into the optimal closure approach, or vice versa.   In particular,
a hybrid methodology might furnish practical approximations of the dissipation potential that are applicable to large amplitudes, 
thereby gaining advantages from both approaches.

\appendix
\section*{Appendix A } 
The closure constants $\gamma_k$ and $\gamma'_k$ occurring in the cost function (\ref{cost-fn-compact})
 are given by the lattice sums  (\ref{gamma}) and (\ref{gamma-prime}).  
 As mentioned in the analysis, it is useful to replace these exact expressions by approximations 
 that capture their $k$-dependence.   

When all the weights $\mu_{p,q}$ determining $\gamma_k$ are equal a common value
and the inverse temperature $\theta$ vanishes, so that $\mu_{p,q} = \mu$ and $\beta_k = \beta$,  
independently of the wavevector indices $p,q,k$,  a simple formula can be deduced
for the dependence of $\gamma_k$ on $k$, over $k \in K_m$, provided that $m \ll n$.     
In this analytically tractable case, the lattice sum (\ref{gamma}) equals the unsymmetrized sum 
\[
\gamma_k \, = \, \frac{\mu^2}{2} \sum_{p+q=k} \frac{ (p \times q)^2}{|q|^2} 
                                  \left( \, \frac{1}{|q|^2} -   \frac{1}{|p|^2}   \right) 
                  \, = \,  \frac{\mu^2}{2} \sum_{q \in K_n} \frac{ (k \times q)^2}{|q|^2} 
                                  \left( \, \frac{1}{|q|^2} -   \frac{1}{|q-k|^2}   \right)  \, ,  
\]
The desired approximation results from replacing this lattice sum over $q \in K_n$
by an integral over $q \in \R^2$.   For the fixed $k$, introduce the polar coordinates, 
$\rho = |q|$, and $\theta$, the angle between $q$ and $k$.    For $\rho \gg |k|$, 
\[
\frac{1}{|q^2|}  - \frac{1}{|q-k|^2} \, = \, 
\frac{1}{\rho^2} - \frac{1}{\rho^2 + |k|^2 - 2 \rho |k| \cos \theta }  \, = \, 
- \frac{ 2 |k| \cos \theta}{\rho^3}  + O\left( \frac{|k|^2}{\rho^4}  \right)  \, .  
\] 
The finite sum over the lattice $K_n$ therefore approaches the convergent infinite sum over 
$\mathbb{Z}^2$ as $n \rightarrow \infty$.   Given that $n \gg m$,  this infinite lattice sum 
is approximated by the corresponding double integral,
\[
\gamma_k \approx \frac{\mu^2}{2} |k|^2 \int_{\frac{1}{2}}^{\infty} \int_0^{2 \pi} 
    \sin^2 \theta \, \left[ \frac{1}{\rho^2} - \frac{1}{\rho^2 + |k|^2 - 2 \rho |k| \cos \theta }  \, \right]
              \rho\,  d\rho \, d\theta \, .  
\]  
Since the inner integral can be evaluated exactly,   
\[
I(\alpha) \, = \, \int_0^{2 \pi} \frac{ \sin^2 \theta \; d \theta}{ 1 + \alpha^2 - 2 \alpha \cos \theta}
\, = \, \left\{   \begin{array} {lll}  \pi & \mbox{ for }  & 0 < \alpha < 1 \\
                                                \pi/\alpha^2   & \mbox{ for }  & \alpha > 1     \end{array}   \right.  ,
\]
it follows that  
\[
  \gamma_k \, \approx \,  \frac{\mu^2}{2} |k|^2 \int_{\frac{1}{2}}^{\infty} 
         \left[ \, \pi - I\left(\frac{|k|}{\rho}\right) \, \right] \,  \frac{d \rho}{\rho}   
    \, = \,  \frac{\pi}{2} \mu^2 \left( \, |k|^2 \log \frac{2}{\sqrt{e}}  |k|  \, + \, \frac{1}{8}   \,  \right)
\]

In light of the conclusion of the Section 6 that the weights can be set to unity, this approximation
suggests a universal model for the $k$-dependence of $\gamma_k$, 
and likewise of $\gamma'_k = 2 \gamma_k$;  namely, 
\begin{equation}  \label{universal-gamma}  
 \gamma_k = C_1 |k|^2 \log C_2 |k| + C_3 \, , \hspace{1cm} \mbox{ with }  \;\; 
    C_1 \approx 1.6, \; C_2 \approx 1.2,  \; C_3 \approx 0.20 \, . 
 \end{equation}    
While the above derivation of this approximate dependence uses $\theta =0$, it remains a good
representation of $\gamma_k$,  for the range of inverse temperature $\theta$ 
encountered in the simulations considered in this paper, except for the lowest wavenumbers.
In general, the variation of $\gamma_k$ with $\theta$ is very weak.          

  \section*{Appendix B }

 This appendix contains the calculations needed to derive the closed reduced equations (\ref{closed-reduced}) 
 by a perturbation expansion of the Hamilton-Jacobi equation (\ref{HJ}) up to cubic order in the resolved
 variables $a_k$ and $y_k = ( b_k - \beta_k) / \beta_k$.   To simplify the algebra, the approximation (\ref{w-approx}) 
 is imposed throughout this derivation.   Under this simplification, and replacing $b_k=\beta_k (1 + y_k) $ by $y_k$, and 
 correspondingly $\psi_k$ by $\eta_k= \beta_k \psi_k$,  the Hamiltonian (\ref{hamiltonian}) is   
\begin{eqnarray}   \label{ham-approx}
\H(a,\phi,y,\eta) & = & \sum_{k \in K_m^+}   \frac{|\phi_k|^2}{\beta_k (1+y_k) } \, + 
                                 \, \phi_k^* f_k(a) + f_k(a)^* \phi_k   \nonumber   \\
                    & & \;\;\;\;\;\;  + \,    \frac{ (1+y_k)^2 \eta_k^2}{2}    \,  - \, \gamma_k  |a_k|^2 
                                       \, - \,  \frac{  \gamma'_k  \, y_k^2}{ 2 \beta_k (1+y_k) }        \nonumber  \\
                        & \approx &    \sum_{k \in K_m^+}   \frac{|\phi_k|^2}{\beta_k} ( 1 - y_k ) \, 
                        +  \, \phi_k^* f_k(a) + f_k(a)^* \phi_k    \\   \nonumber
                    & & \;\;\;\;\;\;        + \,  \frac{ \eta_k^2}{2} ( 1 + 2 y_k )   
                     \,  - \, \gamma_k  |a_k|^2 
                                       \, - \, \frac{\gamma'_k}{2 \beta_k}  y_k^2 ( 1 - y_k ) \, ,
                                              \nonumber              
\end{eqnarray}
where the second expression is the Taylor approximation of $\H$ up to cubic terms.   

The Hamiltonian ({\ref{ham-approx}) is separable over $k$ except for
the terms involving $f_k(a)$.   This special structure  of the associated Hamilton-Jacobi (\ref{HJ}) 
allows us to seek a solution in the power series form:
\begin{eqnarray}   \label{v-expansion}
v(a,y) & = &  \sum_{k \in K_m^+}  [ \,  M_k  + C_k y_k \, ] |a_k|^2 \, + \, 
                      \frac{1}{2} R_k y_k^2 +  \frac{1}{3} S_k y_k^3   \\ 
  & & \;\;\;\;  + \, 
   \frac{1}{3}  \sum_{k \in K_m} \sum_{p+q=k} N_{p,q} \, a_p a_q a_k^*\, + \, \cdots \, .
                       \nonumber
\end{eqnarray} 
The undetermined coefficients, $M_k, C_k, R_k, S_k$, are real, while the coefficients,
$N_{p,q}$ for $p, q \in K_m$, are complex, satisfying $N_{q,p}=N_{p,q}$ and $N_{-p,-q} = N^*_{p,q}$.
The conjugate variables $\phi_k$ and $\eta_k$ satisfy the derived expansions:
\begin{eqnarray*}
 - \phi_k = \frac{\d v}{\d a_k^*} &=&  M_k a_k + C_k a_k y_k + 
        \sum_{p+q=k}   N_{p,q} a_p a_q  + \cdots   \\
 - \eta_k =  \frac{\d v}{\d y_k} &=& C_k |a_k|^2 + R_k y_k + S_k y_k^2 + \cdots \, ,
\end{eqnarray*} 
Upon substituting these expressions into the Hamilton-Jacobi equation for the approximate
Hamiltonian (\ref{ham-approx}),
and equating like terms in the series, we obtain the following equations
for the undetermined coefficients in the expansion (\ref{v-expansion}):   
\begin{eqnarray*}
M_k^2 - \beta_k\gamma_k   &=& 0  \hspace{7mm}  [ \; |a_k|^2 \mbox{ terms } ]  \\
R_k^2 -  \frac{\gamma'_k }{\beta_k }   &=& 0 \hspace{7mm}  [ \; y_k^2 \mbox{ terms } ]  \\ 
 2 M_k C_k   +  \beta_k R_k C_k -  M_k^2   &=& 0 \hspace{7mm} 
           [ \; |a_k|^2 y_k \mbox{ terms } ]  \\
  R_k S_k +   R_k^2  +  \frac{\gamma'_k}{2 \beta_k}  &=& 0 
                  \hspace{7mm} [ \; y_k^3 \mbox{ terms } ]   \\ 
 ( \frac{M_p}{\beta_p} +  \frac{M_q}{\beta_q} + \frac{M_k}{\beta_k} ) N_{p,q}   
         - c(q,k) M_p - c(k,p) M_q + c(p,q) M_k &=& 0       
      \hspace{7mm}  [ \;    a_p a_q a_k^*  \mbox{ terms } ]  
\end{eqnarray*}  
Solving the first four of these equations yields 
\begin{eqnarray}  \label{coefficients}
M_k = \sqrt{\beta_k \gamma_k} \, , \;\;\;\;\;\;  
R_k = \sqrt{\frac{\gamma'_k}{\beta_k}}  \, , \;\;\;\;\;\;
C_k =    \frac{\sqrt{\beta_k} \, \gamma_k}{2\sqrt{\gamma_k}  + \sqrt{\gamma'_k} } \, , 
  \;\;\;\;\;\;
S_k = - \frac{3}{2} \sqrt{\frac{\gamma'_k}{\beta_k}} \, . \nonumber
\end{eqnarray}
The fifth equation yields the coefficients $N_{p,q}$  in terms of the coefficients $M_k$,
and eventually the formula 
\[ 
N_{p,q} = \frac{ c(q,k) \sqrt{\beta_p \gamma_p }  +    
                      c(k,p)  \sqrt{\beta_q \gamma_q }  - c(p,q) \sqrt{\beta_k \gamma_k }   }
        { \sqrt{ \gamma_p / \beta_p }  + \sqrt{ \gamma_q / \beta_q } + \sqrt{ \gamma_k / \beta_k }  }    
        \hspace{1cm}  [  \mbox{ for } p+q=k \, ]  .  
\]

These coefficients determine $v(a,y)$ up to cubic terms, and consequently the desired reduced equations
of the optimal closure are obtained up to quadratic order in $(a,y)$ via the substitutions
\[
\phi_k = \frac{ \d \L}{\d \dot{a_k^*} } = \beta_k (1 + y_k) \left[ \, \dot{a}_k - f_k(a) \,  \right] \, , \hspace{1cm}
 \eta_k = \frac{\d \L}{\d \dot{y}_k}  =   \frac{\dot{y}_k } { (1 + y_k )^2} \,  .    
\]  
That is,  the closed reduced equations (\ref{c-r-a}) and (\ref{c-r-y}) for 
$a_k$ and $y_k$, respectively, follow upon eliminating $\phi_k$ and $\eta_k$.

 \section*{Appendix C}   

Here we collect some standard conventions and calculations 
for functions of several complex variables used in the body of the paper.   
   
For any smooth  function, $f(z)$, of $n$ complex variables, 
$z =(z_1 , \ldots , z_n) \in \C^n$, with $z_k=x_k + i y_k$, the usual derivatives are defined by 
\[
\frac{\d f}{ \d z_k}  = \half  \left( \,  \frac{\d f }{ \d x_k}  - i \frac{\d f}{ \d y_k}  \, \right) \, , \;\;\;\;\;\;\;\; 
\frac{\d f}{ \d z_k^*}  = \half  \left( \,  \frac{\d f}{ \d x_k}  + i \frac{\d f}{ \d y_k}  \, \right) \, ;
\]
the notation $z_k^* = x_k -i y_k$ is used for complex conjugate.    In terms of these,
the chain rule for the composite function $f(z(t))$, where $t$ is a real variable, is simply
\[
\frac{d }{ d t}  f(z(t)) =  \sum_{k=n}^n 
    \frac{\d f}{ \d z_k} \frac{d z_k}{d t} +  \frac{\d f}{ \d z_k^*} \frac{d z_k^*}{ d t}  \, 
              =   \,   \sum_{k=-n}^n   \frac{\d f}{ \d z_k} \frac{d z_k}{d t}  \, .
\]
The last equality invokes the  convention that $z_{-k} = z_k^*$.   In this notation
Taylor's expansion may be written as 
\[
f(z) = f(0) +  \sum_{k=-n}^n L_k \, z_k  +  
      \frac{1}{2} \sum_{k_1,k_2=-n}^n  M_{k_1 k_2} \,  z_{k_1} z_{k_2}
     +  \frac{1}{6} \sum_{k_1,k_2, k_3=-n}^n  N_{k_1 k_2 k_3} \, z_{k_1} \, z_{k_2}z_{k_3}
                       +  O( |z|^4)  \, ,
\]
with coefficients 
\[
 L_k =   \frac{\d f}{ \d z_k}(0) \, , \;\;\;\; 
 M_{k_1 k_2} =   \frac{\d^2 f}{ \d z_{k_1} \d z_{k_2}}(0)  \, , \;\;\;\;
 N_{k_1 k_2 k_3}  =   \frac{\d^3 f}{ \d z_{k_1} \d z_{k_2} \d z_{k_3}} (0)  \, .  
\]
This expansion is used in the perturbation analysis of the value function in Section 5 and Appendix B.     

The Lagrangian $\L(a,\adot, b ,\bdot)$  defined in Section 3, 
and the Hamiltonian $\H(a,\phi,b,\psi)$ and  Hamilton-Jacobi equation 
in Section 4, involve complex conjugate variables, $a \in \C_m$ and $\phi \in \C^m$, related by  
the complexified Legendre transform.   This conjugacy can be seen to be equivalent to the
usual real Legendre transform as follows.    
In a generic notation, consider a Lagrangian $\L(z,\zdot)$ of a complex state vector, $z = x + i y \in C^n$, 
and define the canonically conjugate vector $\zeta = \xi + i \eta \in C^n$ by
\[
\zeta_k = \frac{\d \L}{ \d \zdot_k^*} \, = \, 
   \half \left(  \frac{\d \L}{ \d \xdot_k^*}  + i  \frac{\d \L}{ \d \ydot_k^*}  \right)
    \hspace{2cm}   (\, k =  1, \ldots ,  n \, ) \, , 
\]
and the associated Hamiltonian $\H(z,\zeta)$ by 
\[
\H(z,\zeta)  \, = \, \sum_{k=1}^n \zeta^*_k \zdot_k + \zdot_k^* \zeta_k \, - \, \L \, .    
\]
Then 
\[
 \xi_k = \frac{\d }{ \d \xdot_k }  \, \frac{\L}{2} \, ,   \;\;\;\;   
  \eta_k = \frac{\d }{ \d \ydot_k }  \, \frac{\L}{2}  \, ,  \;\;\;\;\;\;\;
  \frac{\H}{2} \,  = \,    \sum_{k=1}^n \xi_k \xdot_k  + \eta_k \ydot_k \, - \, \frac{\L}{2} \, . 
\]
Thus, the Legendre transform between $\L$ and $\H$ with $n$ complex degrees of freedom
is identical with the real tranform between $\L/2$ and $\H/2$ with $2n$ real degrees of freedom. 

Similarly, the complex Hamilton-Jacobi equation 
\[
\H \left( z, -\frac{ \d v}{\d z^*} \right) \, = \, 0 \, , 
\]
for a value function, $v(z) \in \R$, is equivalent to its real counterpart, and 
the conjugacy relations between $\zeta$ and $z$ along any extremal path 
are 
\[
\zeta_k = - \frac{ \d v}{ \d z_k^*} \, = \, 
    - \frac{\d}{\d x_k} \frac{v}{2}  \, - \, i  \frac{\d}{\d y_k} \frac{v}{2}  \, .      
\]
Thus, this complex Hamilton-Jacobi theory is identical with the usual
real  Hamilton-Jacobi theory
applied to the Hamiltonian $\H(x,\xi,y,\eta) /2$ and the value function $v(x,y)/2$.  
The irrelevant factor $1/2$ merely scales $\L$.

\section*{Acknowlegments}

In the course of this work the authors benefited from conversations
with Richard Kleeman.    The work reported in this paper was partially supported by the National
Science Foundation under grant DMS-1312576.

\end{document}